# Profunctor Optics
## Modular Data Accessors

Matthew Pickering[a], Jeremy Gibbons[b], and Nicolas Wu[a]

a   University of Bristol
b   University of Oxford

**Abstract**   Data accessors allow one to read and write components of a data structure, such as the fields of a record, the variants of a union, or the elements of a container. These data accessors are collectively known as *optics*; they are fundamental to programs that manipulate complex data. Individual data accessors for simple data structures are easy to write, for example as pairs of 'getter' and 'setter' methods. However, it is not obvious how to combine data accessors, in such a way that data accessors for a compound data structure are composed out of smaller data accessors for the parts of that structure. Generally, one has to write a sequence of statements or declarations that navigate step by step through the data structure, accessing one level at a time—which is to say, data accessors are traditionally not first-class citizens, combinable in their own right.

We present a framework for *modular data access*, in which individual data accessors for simple data structures may be freely combined to obtain more complex data accessors for compound data structures. Data accessors become first-class citizens. The framework is based around the notion of *profunctors*, a flexible generalization of functions. The language features required are higher-order functions ('lambdas' or 'closures'), parametrized types ('generics' or 'abstract types') of higher kind, and some mechanism for separating interfaces from implementations ('abstract classes' or 'modules'). We use Haskell as a vehicle in which to present our constructions, but other languages such as Scala that provide the necessary features should work just as well. We provide implementations of all our constructions, in the form of a literate program: the manuscript file for the paper is also the source code for the program, and the extracted code is available separately for evaluation. We also prove the essential properties, demonstrating that our profunctor-based representations are precisely equivalent to the more familiar concrete representations. Our results should pave the way to simpler ways of writing programs that access the components of compound data structures.

**ACM CCS 2012**

- **Software and its engineering → Abstract data types; Patterns; Polymorphism;**

# The Art, Science, and Engineering of Programming

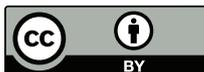

**Profunctor Optics**

# 1 Introduction

Modularity is at the heart of good engineering, since it encourages a separation of concerns whereby solutions to problems can be composed from solutions to subproblems. Compound data structures are inherently modular, and are key to software engineering. A key issue when dealing with compound data structures is in accessing their components—specifically, extracting those components, and modifying them. Since the data structures are compound, we should expect their data accessors to be modular, that is, for data accessors onto compound data structures to be assembled out of data accessors for the components of those structures.

There has been a recent flourishing of work on *lenses* [9] as one such mechanism for data access, bringing programming language techniques to bear on the so-called *view–update problem* [2]. In the original presentation, a lens onto a component of type $A$ within a larger data structure of type $S$ consists of a function $\mathit{view} :: S \to A$ that extracts the component from its context, and a function $\mathit{update} :: A \times S \to S$ that takes a new component of type $A$ and an old data structure of type $S$ and yields a new data structure with the component updated.

We can generalize this presentation, allowing the new component to have a different type $B$ from the old one of type $A$, and the new compound data structure correspondingly to have a different type $T$ from the old one of type $S$. This is a strict generalisation; the earlier definition can be retrieved by specialising the types to $A = B$ and $S = T$. The situation is illustrated in Figure 1(a); the *view* function is a simple arrow $S \to A$, whereas the *update* function has type $B \times S \to T$, so is illustrated by an 'arrow' with two tails and one head. We assemble these functions into a single entity representing the lens as a whole:

    **data** *Lens a b s t* = *Lens* {*view* :: $s \to a$, *update* :: $b \times s \to t$}

This definition declares that the type *Lens*, parametrized by four type variables $a, b, s, t$, has a single constructor, also called *Lens*; a value of that type consists of the constructor applied to a record consisting of two fields, one called *view* of type $s \to a$ and one called *update* of type $b \times s \to t$. (We use an idealization of Haskell as a vehicle for our presentation. See Appendix A for a summary of Haskell notation, and an explanation of those idealizations.)

For example, there is a lens onto the left component of a pair:

    $\pi_1$ :: *Lens a b* $(a \times c)$ $(b \times c)$
    $\pi_1$ = *Lens view update* **where**
       *view* $(x, y)$        = $x$
       *update* $(x', (x, y))$ = $(x', y)$

That is, the value $\pi_1$ is instance of the type *Lens* in which the third and fourth type parameters are instantiated to pair types; the value itself is obtained by applying the constructor *Lens* to the two functions *view* and *update*, whose definitions are provided by the **where** clause. Thus, $\pi_1$ is the lens whose *view* function extracts the first component of a pair, and whose *update* function overwrites that first component.

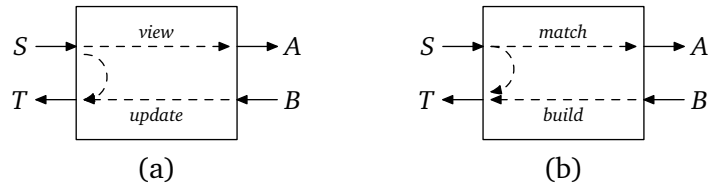

**Figure 1** Illustrations of (a) a lens and (b) a prism. Notice that $update :: B \times S \to T$ (one head and two tails) in (a), but $match :: S \to T + A$ (one tail and two heads) in (b).

In the case of $\pi_1$, the element in focus—the left component of the pair—is directly and separately represented in the compound data structure, as the first component of the pair. The component in focus need not be so explicitly represented. For example, there is also a lens onto the sign of an integer (where, for simplicity, we say that the sign denotes whether the integer is non-negative, rather than being three-valued):

```
sign :: Lens Bool Bool Integer Integer
sign = Lens view update where
    view x       = (x ⩾ 0)
    update (b,x) = if b then abs x else −(abs x)
```

Thus, *sign* is a lens onto a boolean within an integer; the *view* function extracts the sign, and the *update* function enforces a new sign while preserving the absolute value. Note that *sign* is a monomorphic lens, whereas $\pi_1$ was polymorphic: the boolean sign can be replaced only with another boolean, not with a value of a different type.

Analogously to lenses, one could consider compound data structures of several variants. Given a compound data structure of type *S*, one of whose possible variants is of type *A*, one can access that variant via a function $match :: S \to S + A$ that 'downcasts' to an *A* if possible, and yields the original *S* if not; conversely, one may update the data structure via a function $build :: A \to S$ that 'upcasts' a new *A* to the compound type *S*. Such a pair of functions is formally dual to a lens (in the mathematical sense of 'reversing all arrows'), and continuing the optical metaphor has been dubbed a *prism*; prisms are to sum datatypes as lenses are to product datatypes.

More generally, one can again allow the new component to be of a different type *B* from the old one of type *A*, and the new compound structure correspondingly to have a different type *T* from the old one of type *S*. The situation is illustrated in Figure 1(b); this time it is the *build* function of type $B \to T$ that is a simple arrow, whereas *match* has type $S \to T + A$ and performs a case analysis, so is illustrated by an 'arrow' with one tail and two heads. Again, we assemble these two functions together into a record, leading to the following declaration of *Prism* as a four-parameter type, constructed by applying the constructor (also called *Prism*) to a record consisting of two functions *match* and *build*.

```
data Prism a b s t = Prism {match :: s → t + a, build :: b → t}
```

For example, there is a prism onto an optional value, downcasting to that value if present, for which upcasting guarantees its presence:

```
the :: Prism a b (Maybe a) (Maybe b)
the = Prism match build where
   match (Just x) = Right x
   match Nothing  = Left Nothing
   build x        = Just x
```

An optional value of type *Maybe A* is either of the form *Just x* for some $x :: A$, or simply *Nothing*. The *match* field of *the* performs a case analysis on such a value, yielding a result of type *Maybe B + A*, namely *Right x* when the optional *x* is present and *Left Nothing* otherwise. The *build* function simply injects a new value $x :: B$ into the option type *Maybe B*.

Less trivially, there is a prism to provide possible access to a floating-point number as a whole number, 'downcasting' a *Double* to an *Integer* whenever the fractional part is zero.

```
whole :: Prism Integer Integer Double Double
whole = Prism match build where
   match x
      | f == 0    = Right n
      | otherwise = Left x
      where (n, f) = properFraction x
   build = fromIntegral
```

The *fromIntegral* function converts from *Integer* to *Double*, while *properFraction x* returns a pair $(n, f)$ where $n :: Integer$ is the integer part of *x* and $f :: Double$ is the fractional part. (Of course, there are the usual caveats about floating-point precision.)

For any given kind of compound data structure, it is usually straightforward to provide such accessors. However, when it comes to composing data structures out of parts, the data accessors do not compose conveniently. For example, there is a lens onto the leftmost component *A* of a nested pair $(A \times B) \times C$; but it is very clumsy to write that lens in terms of the existing lens $\pi_1$ for non-nested pairs:

```
π₁₁ :: Lens a b ((a × c) × d) ((b × c) × d)
π₁₁ = Lens view update where
   Lens v u       = π₁
   view           = v · v
   update (x′, xyz) = u (xy′, xyz) where
                      xy  = v xyz
                      xy′ = u (x′, xy)
```

(here, the first local definition matches the pattern *Lens v u* against $\pi_1$, thereby binding *v* and *u* to the two fields of $\pi_1$). In fact, for the update method it is clearer to resort instead to first principles, defining $update\ (x', ((x, y), z)) = ((x', y), z)$; this points to a failure of modularity in our abstraction.

The situation is even worse for heterogeneous composite data structures. For example, we might want to access the *A* component of a compound type $(1 + A) \times B$ built using both sums and products. We might further hope to be able to construct this accessor onto the optional left-hand component by composing the prism *the* onto an optional value and the lens $\pi_1$ onto a left-hand component. However, the composite accessor is not a lens, because it cannot guarantee a view as an *A*; neither is it a prism, because it cannot build the composite data structure from an *A* alone. We cannot even express this combination; our universe of accessors is not closed under the usual operations for composing data. The abstraction is clearly broken.

Lenses and prisms, and some other variations that we will see in Section 2, have collectively been called *optics*. In this paper, we present a different representation of optics, which fixes the broken abstraction for lenses, prisms, and the other optics. The representation is based on the notion of *profunctors*, a generalization of functions; we introduce and explain the ideas as we go along. We call this new representation *profunctor optics*. It provides accessors for composite data that are trivially composed from accessors for the parts of the data, using ordinary function composition: for example, lenses become easily combinable with lenses, and lenses become combinable at all with prisms—hence *modular data accessors*. Moreover, the profunctor representation reveals a lattice structure among varieties of optics—structure that remains hidden with the concrete representations.

The constructions we present are not in fact new; they are Haskell folklore, having been introduced by others in the form of Internet Relay Chat comments, blog posts, sketches of libraries, and so on. But they deserve to be better known; our main contribution is to write these constructions up in a clear and consistent manner.

The remainder of this paper is structured as follows. In Section 2, we introduce a common specialization of lenses and prisms (called *adapters*) and a common generalization (called *traversals*). Section 3 introduces the notion of *profunctor*, the main technical device on which we depend. Section 4 revisits the four varieties of optic with new representations in terms of profunctors. Section 5 shows how the profunctor representation supports modular construction of accessors for compound data structures in ways that the concrete representations do not. Section 6 summarizes prior and related work, and Section 7 concludes.

The paper itself is a literate script; all the code is embedded in the paper, has been explicitly type-checked, and is available for experimentation [33]. We do not make essential use of any fancy Haskell features; all that is really needed are higher-order functions, higher-kinded parametrized types, and some mechanism for separating interfaces from implementations, all of which are available in other languages. For the benefit of non-Haskellers, Appendix A summarises the Haskell notation and standard functions that we use; Appendix B sketches an alternative implementation in Scala; and Appendix C formally states and proves equivalences between the concrete and profunctor optic representations.

## 2  Optics, concretely

We have already seen two distinct varieties of optic, namely lenses and prisms. It turns out that they have a common specialization, which we call *adapters*, and a common generalization, which we call *traversals*, both of which we introduce in this section.

**Profunctor Optics**

## 2.1 Adapters

When the component being viewed through a lens is actually the whole of the structure, then the lens is essentially a pair of functions of types $S \to A$ and $B \to T$; there is no virtue in the update function taking the old $S$ value as an argument as well, because this will be completely overwritten. Dually, when the variant being accessed through a prism is actually the sole variant, the prism is again essentially a pair of functions of types $S \to A$ and $B \to T$; there is no virtue in the matching function having a fallback option, because the match will always succeed. This is illustrated in Figure 2. We introduce an abstraction for such a pair of functions, which we call an *adapter*.

> **data** *Adapter a b s t* = *Adapter* {*from* :: $s \to a$, *to* :: $b \to t$}

It is often the case that *from* and *to* are in some sense each other's inverses; but we will not attempt to enforce that property, nor shall we depend on it.

Although adapters look like rather trivial data accessors, they are very useful as 'plumbing' combinators, converting between representations. For example, we introduced earlier the composite optic $\pi_{11}$ to access the $A$ component buried within a nested $(A \times B) \times C$ tuple. Now suppose that we want to access the $A$ component in a different but isomorphic tuple type $A \times B \times C$ as well. We do not need a separate lens in order to allow this; it suffices to combine $\pi_{11}$ with the isomorphism

> *flatten* :: *Adapter* $(a \times b \times c)$ $(a' \times b' \times c')$ $((a \times b) \times c)$ $((a' \times b') \times c')$
> *flatten* = *Adapter from to* **where**
>   *from* $((x,y),z) = (x,y,z)$
>   *to* $(x,y,z)\ \ \ \ = ((x,y),z)$

which serves as an adapter between the two tuple types.

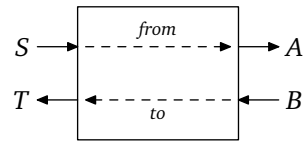

**Figure 2** An adapter.

## 2.2 Traversal

A *traversable* datatype is a container datatype (such as lists, or trees), in which the data structures a finite number of elements, and an ordering on the positions of those elements. Given such a traversable data structure, one can *traverse* it, visiting each of the elements in turn, in the given order. When the container is polymorphic, one may vary the type of the elements in the process; for example, turning a tree of integers into a tree of characters. Moreover, because the ordering on positions is explicit, one may safely apply an *effectful* operation to each element, for example performing I/O or manipulating some mutable variable; the traversal of the whole structure sequences the effects arising from the elements in an order determined by the positions of those elements [12, 23].

In a pure language such as Haskell, we express a class of effects as a datatype of effectful computations. The best known example of such a datatype is *monads* [39].

It turns out that we do not need the full expressive power of monads for traversals; the more restrictive abstraction of *applicative functors* [23] suffices. The interfaces modelling functors and applicative functors are represented using type classes:

```
class Functor f where
   fmap :: (a → b) → f a → f b
class Functor f ⇒ Applicative f where
   pure :: a → f a
   (⟨∗⟩) :: f (a → b) → f a → f b
```

These declarations introduce two classes *Functor* and *Applicative* of operations on types, with the latter a subclass of the former. For $F$ to be a functor, one has to provide a function *fmap* of the declared type; one can think of the type $F\ A$ denoting a certain kind of 'containers of $A$s', such as lists, and of *fmap f* as applying $f$ to each element of such a container. Similarly, one can think of applicative functor $F$ as representing a certain class of effects, and the type $F\ A$ as the type of 'computations that may have effects of type $F$ when run, and will yield a result of type $A$'; ordinary functions of type $A \to F\ B$ can be thought of as 'effectful functions' from $A$ to $B$, having effects modelled by $F$. The *pure* operation lifts a plain value to a trivial computation that actually has no effects and simply returns the given value; alternatively, one can think of *pure* itself as an effectful version of the identity function. The operator ⟨∗⟩ acts to combine computations: if $m :: F\ (A \to B)$ is a computation that effectfully returns an $A \to B$ function, and $n :: F\ A$ similarly a computation that effectfully returns an $A$, then $m\ ⟨∗⟩\ n$ is the composite computation that runs $m$ to get an $A \to B$ function and runs $n$ to get an $A$ argument, then applies the function to the argument to return a $B$ result overall, incurring the effects of both $m$ and $n$.

As an example of a computation type, consider stateful computations represented as state-transforming functions:

```
data State s a = State {run :: s → a × s}
```

so that the computation that increments an integer counter and returns a given boolean value is captured by the definition

```
inc :: Bool → State Integer Bool
inc b = State (λn → (b, n + 1))
```

For any state type $S$, the type *State S* is an applicative functor:

```
instance Functor (State s) where
   fmap f m = State (λs → let (x, s′) = run m s in (f x, s′))
instance Applicative (State s) where
   pure x   = State (λs → (x, s))
   m ⟨∗⟩ n  = State (λs → let (f, s′)  = run m s
                              (x, s″) = run n s′
                          in (f x, s″))
```

in which mapping applies a function to the returned result, a pure computation leaves the state unchanged, and sequential composition threads the state through the first then through the second computation.

Now, for an applicative functor $F$, a traversal takes an effectful operation of type $A \to F\,B$ on the elements of a container, and lifts this to an effectful computation of type $S \to F\,T$ over the whole container, applying the operation to each element in turn. We say that container type $S$ with elements of type $A$ is *traversable* when there exist types $B, T$ and a traversal function of type $(A \to F\,B) \to (S \to F\,T)$ for each applicative functor $F$ (which should satisfy some laws [12], not needed here).

For example, consider the datatype

    **data** *Tree a* = *Empty* | *Node* (*Tree a*) *a* (*Tree a*)

of internally labelled binary trees, in which constructor *Empty* represents the empty tree and *Node t x u* the non-empty tree with root labelled $x$ and children $t, u$. This datatype is traversable; one of several possible orders of traversal is in-order:

    *inorder* :: *Applicative f* $\Rightarrow$ $(a \to f\,b) \to$ *Tree a* $\to f$ (*Tree b*)
    *inorder m Empty*         = *pure Empty*
    *inorder m* (*Node t x u*) = ((*pure Node* ⟨∗⟩ *inorder m t*) ⟨∗⟩ *m x*) ⟨∗⟩ *inorder m u*

which visits the root $x$ after visiting the left child $t$ and before visiting the right child $u$. (The sequencing operation ⟨∗⟩ is declared to associate to the left, like function application does, so the parentheses on the right-hand side of the second equation are redundant; we will henceforth omit them.) Thus, effectfully traversing each of the elements of the empty tree is pure, yielding simply the empty tree; and traversing a non-empty tree yields the (pure) assembly of the results of recursively traversing its left child, operating on the root label, and recursively traversing the right child, with the effects occurring in that order. For example, the computation

    *countOdd* :: *Integer* $\to$ *State Integer Bool*
    *countOdd n* = **if** *even n* **then** *pure False* **else** *inc True*

increments a counter when the argument is odd and leaves it unchanged when the argument is even, and returns the parity of that argument; and so

    *inorder countOdd* :: *Tree Integer* $\to$ *State Integer* (*Tree Bool*)

performs an in-order traversal of a tree of integers, counting the odd ones, and returning their parities.

### 2.3 Traversals as concrete optics

Traversal can be seen as a generalisation of lenses and of prisms, providing access not just to a single component within a whole structure but onto an entire sequence of such components. Indeed, the type $(A \to F\,B) \to (S \to F\,T)$ of witnesses to traversability of the container type $S$ is almost equivalent to a pair of functions *contents* :: $S \to A^n$ and $fill :: S \times B^n \to T$, for some $n$ being the number of elements in the container. The idea is that *contents* yields the sequence of elements in the container, in the order specified by the traversal, and *fill* takes an old container and a new sequence of elements and updates the old container by replacing each of the elements with a new one. Roughly speaking, for singleton containers ($n = 1$) this specialises both to lenses and to prisms. However, a factorization into two functions *contents* and *fill* is not quite right, because the appropriate value of the exponent $n$ depends on the particular container in $S$, and must match for applications of *contents* and *fill*: one can in general only refill a container with precisely the same number of elements as it originally contained. However, the dependence can be captured by tupling together the two functions and using a common existentially quantified length: the traversable type $S$ is equivalent to $\exists n \,.\, A^n \times (B^n \to T)$. This fact is not obvious, but is well established [3, 14].

We can capture the result of that tupled pair of functions via the following datatype:

**data** $FunList\ a\ b\ t = Done\ t \mid More\ a\ (FunList\ a\ b\ (b \to t))$

This datatype was introduced by van Laarhoven [19]. It is a so-called *nested* datatype [4], because in the *More* case a larger value of type $FunList\ A\ B\ T$ is constructed not from smaller values of the same type, but from a value of a different type $FunList\ A\ B\ (B \to T)$. One may verify inductively that $FunList\ A\ B\ T$ is isomorphic to $\exists n \,.\, A^n \times (B^n \to T)$: the *Done* case consists of simply a $T$, corresponding to $n = 0$; and the *More* case consists of an $A$ and an $A^n \times (B^n \to (B \to T))$ for some $n$, and by isomorphisms of products and function spaces we have $A \times (A^n \times (B^n \to (B \to T))) \simeq A^{n+1} \times (B^{n+1} \to T)$.

The isomorphism between $FunList\ A\ B\ T$ and $T + (A \times (FunList\ A\ B\ (B \to T)))$ is witnessed by the following two functions:

```
out :: FunList a b t → t + (a, FunList a b (b → t))
out (Done t)   = Left t
out (More x l) = Right (x, l)

inn :: t + (a, FunList a b (b → t)) → FunList a b t
inn (Left t)        = Done t
inn (Right (x, l))  = More x l
```

Now, a traversal function of type $(A \to F\ B) \to (S \to F\ T)$ for each applicative functor $F$ yields an isomorphism $S \simeq FunList\ A\ B\ T$. In order to construct the transformation from $S$ to $FunList\ A\ B\ T$ using such a traversal function, we require $FunList\ A\ B$ to be an applicative functor:

```
instance Functor (FunList a b) where
  fmap f (Done t)   = Done (f t)
  fmap f (More x l) = More x (fmap (f·) l)

instance Applicative (FunList a b) where
  pure            = Done
  Done f ⟨∗⟩ l'   = fmap f l'
  More x l ⟨∗⟩ l' = More x (fmap flip l ⟨∗⟩ l')
```

The actual definitions may appear obscure, but in essence they make *FunList*s a kind of sequence, with three operations corresponding to mapping, the empty sequence, and concatenation. We also require an operation of type $A \to \mathit{FunList}\ A\ B\ B$ on elements, which we call *single* as it parcels up an element as a singleton *FunList*:

    *single* :: $a \to \mathit{FunList}\ a\ b\ b$
    *single x* = *More x* (*Done id*)

We can use *single* as the body of a traversal, instantiating the applicative functor *F* to *FunList A B*. This traversal will construct a singleton *FunList* for each element of a container, then concatenate the singletons into one long *FunList*. In particular, this gives *t single* :: $S \to \mathit{FunList}\ A\ B\ T$ as one half of the isomorphism $S \simeq \mathit{FunList}\ A\ B\ T$. Conversely, we can retrieve the traversable container from the *FunList*:

    *fuse* :: $\mathit{FunList}\ b\ b\ t \to t$
    *fuse* (*Done t*)   = *t*
    *fuse* (*More x l*) = *fuse l x*

This motivates the following definition of concrete traversals:

    **data** *Traversal a b s t* = *Traversal* {*extract* :: $s \to \mathit{FunList}\ a\ b\ t$}

The situation is illustrated in Figure 3. The upper collection of left-to-right arrows represents the *contents* aspect, extracting the sequence of *A* elements from a container; the lower collection of right-to-left arrows represents the *fill* aspect, generating a new container from a fresh sequence of *B* elements. However, it is more precise to think of these as one combined arrow *extract* from *S*, generating both the *A*s and the mapping back from the *B*s to the *T*.

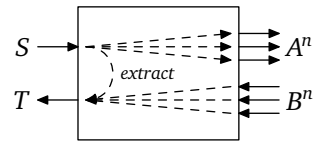

**Figure 3** A traversal. Sole method *extract* has type $S \to A^n \times (B^n \to T)$.

As another example, *inorder single* :: *Tree a* $\to \mathit{FunList}\ a\ b\ (\mathit{Tree}\ b)$ extracts the in-order sequence of elements from a tree, and moreover provides a mechanism to refill the tree with a new sequence of elements. This type matches the payload of a concrete traversal; so we can define concrete in-order traversal of a tree by:

    *inorderC* :: *Traversal a b* (*Tree a*) (*Tree b*)
    *inorderC* = *Traversal* (*inorder single*)

## 3 Profunctors

The key to the design of a modular abstraction for data accessors is to identify what they have in common. Any data accessor for a component of a data structure is 'function-like', in the sense that reading 'consumes' the component from the data structure and writing 'produces' an updated component to put back into the data structure. The type structure of such function-like things—henceforth *transformers*—is

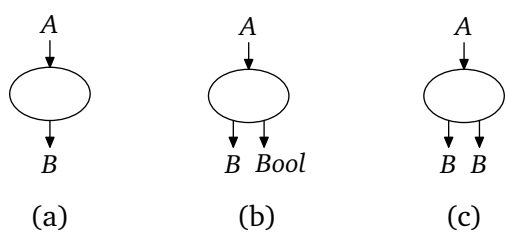

**Figure 4** A transformer of type *P A B* 'consumes *A*s and produces *B*s'.

technically known as a *profunctor*. Profunctors can be represented by the following type class:

**class** *Profunctor p* **where**
    *dimap* :: $(a' \to a) \to (b \to b') \to p\ a\ b \to p\ a'\ b'$

This says that the two-place operation on types *P* is a *Profunctor* if there is a suitable definition of the function *dimap* with the given type. Think of *P A B* as a type of 'transformers that consume *A*s and produce *B*s', with different instantiations of *P* corresponding to different notions of 'function-like', as illustrated in Figure 4. One can think of *dimap f g h* as 'wrapping' the transformer *h* in a preprocessor *f* and a postprocessor *g*. The crucial point is that a transformer is *covariant* in what it produces, but *contravariant* in what it consumes; hence the reversal of the arrow ($a' \to a$ rather than $a \to a'$) in the type of the preprocessor, the first argument of *dimap*. The term 'profunctor' comes from category theory, although much of the categorical structure gets lost in translation.

Instances of the *Profunctor* class should satisfy two laws about the interaction between *dimap* and function composition:

*dimap id id* $\quad\quad\quad$ = *id*
*dimap* $(f' \cdot f)\ (g \cdot g')$ = *dimap f g* $\cdot$ *dimap f' g'*

Note again the contravariance in the preprocessor argument in the second law.

The canonical example of transformers is, of course, functions themselves; and indeed, the function arrow $\to$ on types, for which $(\to)\ A\ B = A \to B$, is an instance:

**instance** *Profunctor* $(\to)$ **where**
    *dimap f g h* = $g \cdot h \cdot f$

The reader should see that the contravariance in the first argument is necessary. It is instructive to verify for oneself that the definition of *dimap* for function types does indeed satisfy the two profunctor laws.

Plain functions are, of course, not the only instantiation of the abstraction—if they were, the abstraction would not be very useful. Functions that return a result together with a boolean flag are another instance, as illustrated in Figure 4(b). So are functions that return a pair of results, as illustrated in Figure 4(c). This pattern generalizes to functions of the form $A \to F\ B$ for some functor *F*:

**data** *UpStar f a b* = *UpStar* {*unUpStar* :: $a \to f\ b$}

**Profunctor Optics**

Any functor $F$ lifts in this manner to a profunctor:

   **instance** *Functor f* $\Rightarrow$ *Profunctor* (*UpStar f*) **where**
     *dimap f g* (*UpStar h*) $=$ *UpStar* (*fmap g* $\cdot$ *h* $\cdot$ *f*)

(Indeed, the construction dualizes, to functions of the form $F\,A \to B$, and the two constructions may be combined for functions of the form $F\,A \to G\,B$; but we do not need that generality for this paper.)

We will, however, have need of three refinements of the notion of profunctor, concerning its interaction with product and sum types. For the first, we say that a profunctor is *cartesian* if, informally, it can pass around some additional context in the form of a pair. This is represented by an additional method *first* that lifts a transformer of type $P\,A\,B$ to one of type $P\,(A \times C)\,(B \times C)$ for any type $C$, passing through an additional contextual value of type $C$:

   **class** *Profunctor p* $\Rightarrow$ *Cartesian p* **where**
     *first*   :: $p\,a\,b \to p\,(a \times c)\,(b \times c)$
     *second* :: $p\,a\,b \to p\,(c \times a)\,(c \times b)$

For each instance $P$, the method *first* should satisfy two additional laws, concerning coherence with product and the unit type:

   *dimap runit runit$'$ h*                $=$ *first h*
   *dimap assoc assoc$'$* (*first* (*first h*)) $=$ *first h*

(and symmetrically for *second*), where *runit* :: $a \times 1 \to a$ and *runit$'$* :: $a \to a \times 1$ are witnesses to the unit type being a right unit of the cartesian product, and *assoc* :: $a \times (b \times c) \to (a \times b) \times c$ and *assoc$'$* :: $(a \times b) \times c \to a \times (b \times c)$ are witnesses to the associativity of product. (Note the typing in the unit law, which instantiates $C$ to 1: instead of passing around trivial additional context, one may discard it then recreate it.) To be precise, one might call such profunctors *cartesianly strong*, because *first* acts as a categorical 'strength' with respect to cartesian product; we abbreviate this more precise term to simply 'cartesian'. The function arrow is obviously cartesian:

   **instance** *Cartesian* ($\to$) **where**
     *first h*   $=$ *cross h id*
     *second h* $=$ *cross id h*

(where *cross f g* $(x,y) = (f\,x, g\,y)$ applies two functions to a pair of arguments). So too are functions with structured results, as captured by *UpStar*:

   **instance** *Functor f* $\Rightarrow$ *Cartesian* (*UpStar f*) **where**
     *first* (*UpStar unUpStar*)    $=$ *UpStar* (*rstrength* $\cdot$ *cross unUpStar id*)
     *second* (*UpStar unUpStar*) $=$ *UpStar* (*lstrength* $\cdot$ *cross id unUpStar*)

where the so-called 'right strength'

   *rstrength* :: *Functor f* $\Rightarrow$ $((f\,a) \times b) \to f\,(a \times b)$
   *rstrength* $(fx, y) =$ *fmap* $(, y)\,fx$

distributes copies of a *B*-value over an *F A* structure, and symmetrically for left strength—here, the function (,*y*) takes *x* to (*x*,*y*). But it is not always so obvious how to thread the contextual values through a profunctor. In particular, there is no general construction for the dual case of functions with structured arguments. For example, when *F* is the functor *Pair* yielding pairs of elements, the dual case entails putting together a function of type *Pair A* → *B* with a *Pair* (*A* × *C*) to make a *B* × *C*; there are two input *C*s from which to choose the output, with neither being canonical. Worse, when *F* = *Maybe*, there is not necessarily a *C* in the input at all.

Similarly, there is a refinement of profunctors that can be lifted to act on sum types:

**class** *Profunctor p* ⇒ *Cocartesian p* **where**
   *left*   :: *p a b* → *p* (*a* + *c*) (*b* + *c*)
   *right* :: *p a b* → *p* (*c* + *a*) (*c* + *b*)

Informally, if *h* :: *P A B* is a transformer of *A*s into *B*s, then *left h* :: *P* (*A* + *C*) (*B* + *C*) acts on the *A*s in a sum type *A* + *C*, turning them into *B*s and leaving the *C*s alone.

For each instance *P*, the method *left* should satisfy two additional laws, concerning coherence with sum and the empty type:

   *dimap rzero rzero′ h*                = *left h*
   *dimap coassoc′ coassoc* (*left* (*left h*)) = *left h*

(and symmetrically for *right*), where *rzero* :: *a*+0 → *a* and *rzero′* :: *a* → *a*+0 are witnesses to the empty type 0 being a right unit of sum, and *coassoc* :: *a* + (*b* + *c*) → (*a* + *b*) + *c* and *coassoc′* :: (*a* + *b*) + *c* → *a* + (*b* + *c*) are witnesses to the associativity of sum. (Again, note the typing of the zero law, instantiating *C* = 0: instead of lifting to a trivial sum, one may discard and then recreate the trivial missing information.) To be precise, one might call such profunctors *co-cartesianly strong*, because the methods act as left and right strengths for the co-Cartesian structure; but we will stick with the abbreviation 'co-cartesian'.

The function arrow is obviously co-cartesian:

**instance** *Cocartesian* (→) **where**
   *left h*   = *plus h id*
   *right h* = *plus id h*

(where *plus f g* takes *Left x* to *Left* (*f x*) and *Right y* to *Right* (*g y*)). So too are functions with structured results, provided that there is an injection *A* → *F A* of pure values into that structure. For convenience, we capture that requirement here as the method *pure* of the type class *Applicative*, since we have introduced this already:

**instance** *Applicative f* ⇒ *Cocartesian* (*UpStar f*) **where**
   *left* (*UpStar unUpStar*)   = *UpStar* (*either* (*fmap Left* · *unUpStar*) (*pure* · *Right*))
   *right* (*UpStar unUpStar*) = *UpStar* (*either* (*pure* · *Left*) (*fmap Right* · *unUpStar*))

However, that constraint is stronger than necessary, because we do not need here the ⟨∗⟩ method of the *Applicative* class. (Again, there is no similar construction for functions with structured arguments.)

**Profunctor Optics**

The third refinement is a class of profunctors that support a form of parallel composition (in the sense of 'independent' rather than 'concurrent'):

**class** *Profunctor p* ⇒ *Monoidal p* **where**
  *par*    :: *p a b* → *p c d* → *p* (*a* × *c*) (*b* × *d*)
  *empty* :: *p* 1 1

Informally, if $h :: P\,A\,B$ and $k :: P\,C\,D$ are transformers of $A$s into $B$s and of $C$s into $D$s, respectively, then *par h k* transforms $A \times C$ pairs into $B \times D$ pairs by acting independently on each component of the pair; and *empty* is a trivial transformer of unit values into unit values.

For each *Monoidal* instance $P$, the two operations *par* and *empty* should satisfy some laws concerning coherence with the product structure: they should form a monoid, up to monoidal isomorphisms on the value types:

*dimap assoc assoc'* (*par* (*par h j*) *k*) = *par h* (*par j k*)
*dimap runit runit' h*             = *par h empty*
*dimap lunit lunit' h*              = *par empty h*

where $lunit :: 1 \times a \to a$ and $lunit' :: a \to 1 \times a$ are the witnesses to the unit type being the left as well as the right unit of cartesian product.

The function arrow is obviously monoidal:

**instance** *Monoidal* (→) **where**
  *par*   = *cross*
  *empty* = *id*

For functions with structured results (that is, of the form $A \to F\,B$) to be monoidal, it is necessary to be able to 'zip' together two $F$-structures. This can be done when $F$ is an applicative functor:

**instance** *Applicative f* ⇒ *Monoidal* (*UpStar f*) **where**
  *empty*  = *UpStar pure*
  *par h k* = *UpStar* (*pair* (*unUpStar h*) (*unUpStar k*))

where for the definition of *par* we make use of the lifting

*pair* :: *Applicative f* ⇒ (*a* → *f b*) → (*c* → *f d*) → (*a*, *c*) → *f* (*b*, *d*)
*pair h k* (*x*, *y*) = *pure* (,) ⊛ *h x* ⊛ *k y*

to applicative functors of the pairing function (,) defined by (,) *x y* = (*x*, *y*).

## 4   Optics in terms of profunctors

Plain data accessors might be modelled simply as transformers, values of some type that is an instance of the *Profunctor* type class as discussed above. However, such a Matthew Pickering, Jeremy Gibbons, and Nicolas Wumodel will not address the problems of compositionality that motivated us in the first place. Instead, we represent data accessors as *mappings between transformers*:

**type** *Optic p a b s t* = *p a b* → *p s t*

Informally, when *S* is a composite type with some component of type *A*, and *T* similarly a composite type in which that component has type *B*, and *P* is some notion of transformer, then we can think of a data accessor of type *Optic P A B S T* as lifting a component transformer of type *P A B* to a whole-structure transformer of type *P S T*. We will retrieve equivalents of our original definitions of lens, prism, and so on by placing various constraints on *P*, starting with requiring *P* to be a *Profunctor*. Crucially, different varieties of optic all now have the same form—in particular, they are all simply functions—and so they will compose straightforwardly; they may involve different constraints on *P*, but those constraints simply conjoin.

The situation is somewhat analogous to that with real, imaginary, and complex numbers. The concrete representations of optics are like having real numbers and imaginary numbers but not arbitrary complex numbers. One can combine two real numbers using addition to make a third, and combine two imaginary numbers to make a third; but one cannot combine a real number and an imaginary number with addition, because the result is in general neither real nor imaginary. But once one invents complex numbers, now arbitrary combinations by addition are expressible. Moreover, the complex numbers embed faithful representations of the real numbers and of the imaginary numbers; the space of complex numbers is strictly richer than the union of the spaces of real and of imaginary numbers. (The analogy only goes so far. Composition of concrete lenses is not easily defined, and only becomes straightforward in the profunctor representation. It is as if addition of real numbers becomes more easily expressed by passage through the complex numbers.)

### 4.1 Profunctor adapters

Recall the concrete representation of adapters from Section 2.1:

**data** *Adapter a b s t* = *Adapter* {*from* :: *s* → *a*, *to* :: *b* → *t*}

The two methods *from* and *to* of an *Adapter A B S T* do not generally compose, specifically when types *A* and *B* differ. However, if we could somehow transform *A*s into *B*s, then we could make the two methods fit together; and moreover, we would then be able to transform *S*s into *T*s in the same way. Which is to say, there is an obvious mapping that takes an *Adapter A B S T* and a *P A B* and yields a *P S T*, provided that *P* is a profunctor. This motivates the following datatype:

**type** *AdapterP a b s t* = ∀*p* . *Profunctor p* ⇒ *Optic p a b s t*

That is, an optic of type *AdapterP A B S T* is simply a function from *P A B* to *P S T* that works polymorphically in the profunctor type *P*. It will turn out, somewhat surprisingly, that *AdapterP A B S T* is precisely equivalent to *Adapter A B S T* (see Appendix C for the proof); we are therefore justified in using *AdapterP* as a profunctor representation of adapters.

7-**15**

**Profunctor Optics**

The translations between the two representations are not difficult to construct. We have already hinted at the translation from the concrete representation *Adapter* to the profunctor representation *AdapterP*:

>   *adapterC2P* :: *Adapter a b s t* → *AdapterP a b s t*
>   *adapterC2P* (*Adapter o i*) = *dimap o i*

This definition repays a little contemplation: given functions $o :: S \to A$ and $i :: B \to T$, then *dimap o i* has type $P\ A\ B \to P\ S\ T$ for any profunctor *P*, as required.

The translation in the opposite direction takes a little more effort: what can we do with an *l* of type *AdapterP A B S T*? This function has type $P\ A\ B \to P\ S\ T$ for arbitrary profunctor *P*; if we are to use it somehow to construct an *Adapter A B S T*, then it had better be the case that *Adapter A B* is a profunctor, a suitable instantiation for *P*. Happily, this is the case:

>   **instance** *Profunctor* (*Adapter a b*) **where**
>     *dimap f g* (*Adapter o i*) = *Adapter* (*o* · *f*) (*g* · *i*)

Informally, this *dimap* wraps the pair of functions $o :: S \to A$ and $i :: B \to T$ in a preprocessor $f :: S' \to S$ and postprocessor $g :: T \to T'$:

$$S' \xrightarrow{f} S \xrightarrow{o} A \qquad B \xrightarrow{i} T \xrightarrow{g} T'$$

to yield a pair of functions of types $S' \to A$ and $B \to T'$. It is straightforward to check that this definition satisfies the two profunctor laws.

Now, we construct the trivial concrete adapter *Adapter id id* of type *Adapter A B A B*, and use the profunctor adapter to lift that to the desired concrete adapter:

>   *adapterP2C* :: *AdapterP a b s t* → *Adapter a b s t*
>   *adapterP2C l* = *l* (*Adapter id id*)

Again, it is a worthwhile exercise to verify the types: function *l* is applicable at arbitrary profunctors *P*, but we use it here only for the specific profunctor *P* = *Adapter A B*; then *l* transforms a *P A B* into a *P S T*.

Note the essential use of profunctors in the translation. For *adapterC2P*, it is tempting to pick a simpler translation: given an *Adapter A B S T*, which is a pair of functions of types $S \to A$ and $B \to T$, and another function of type $A \to B$, then one can construct a function of type $S \to T$; that is, one can translate from *Adapter A B S T* to the pure function type $(A \to B) \to (S \to T)$. But this translation loses information, because there is no obvious translation back from here to the profunctor representation—in particular, it provides no way of constructing anything other than a pure function.

The proof that *adapterC2P* and *adapterP2C* are each other's inverses, and hence that *Adapter A B S T* and *AdapterP A B S T* are equivalent, can be found in Appendix C. For the remaining varieties of optic, we present the constructions and discussion in a bit less detail.

### 4.2 Profunctor lenses

Recall the concrete representation of lenses from Section 1:

**data** *Lens a b s t* = *Lens* {*view* :: *s* → *a*, *update* :: *b* × *s* → *t*}

The occurrence of the product type in the argument of *update* suggests that the analogue of *Lens* will have something to do with cartesian profunctors. Indeed, we define the profunctor representation of lenses as follows:

**type** *LensP a b s t* = ∀*p* . *Cartesian p* ⇒ *Optic p a b s t*

That is, a profunctor lens *LensP A B S T* lifts a transformer on components *P A B* to a transformer on structures *P S T*, for arbitrary cartesian profunctor *P*.

Concrete lenses are themselves cartesian profunctors:

**instance** *Profunctor* (*Lens a b*) **where**
   *dimap f g* (*Lens v u*) = *Lens* (*v* · *f*) (*g* · *u* · *cross id f*)
**instance** *Cartesian* (*Lens a b*) **where**
   *first* (*Lens v u*)    = *Lens* (*v* · *fst*) (*fork* (*u* · *cross id fst*) (*snd* · *snd*))
   *second* (*Lens v u*) = *Lens* (*v* · *snd*) (*fork* (*fst* · *snd*) (*u* · *cross id snd*))

(where *fork f g x* = (*f x*, *g x*) applies two functions to a common argument to return a pair). The translations back and forth make crucial use of the lifting to products. For the translation from the concrete representation to the profunctor representation, we need to translate a concrete lens *Lens v u* :: *Lens A B S T* to a profunctor optic, which is a function of type *Optic P A B S T* that has to work for an arbitrary cartesian profunctor *P*. In other words, given *v* :: *S* → *A* and *u* :: *B* × *S* → *T* and a transformer *h* :: *P A B* for some cartesian profunctor *P*, we have to construct another transformer of type *P S T*. Now, *first h* has type *P* (*A* × *C*) (*B* × *C*) for any type *C*, and in particular, for *C* = *S*. Then it suffices to wrap this transformer in a preprocessor *S* → *A* × *S* and a postprocessor *B* × *S* → *T*, both of which are easy to construct:

   *lensC2P* :: *Lens a b s t* → *LensP a b s t*
   *lensC2P* (*Lens v u*) = *dimap* (*fork v id*) *u* · *first*

For the translation in the opposite direction, we use the same approach as for adapters. We are given the profunctor lens, *l* :: *Optic P A B S T*, which will work for arbitrary cartesian profunctor *P*; we have to construct a concrete lens of type *Lens A B S T*. We note that *Lens A B* is itself a cartesian profunctor, so *l* is applicable at the type *P* = *Lens A B*. We therefore construct the trivial concrete lens *Lens id fst* :: *Lens A B A B*, and lift it using *l* to a *Lens A B S T* as required:

   *lensP2C* :: *LensP a b s t* → *Lens a b s t*
   *lensP2C l* = *l* (*Lens id fst*)

These definitions may seem somewhat mysterious; and indeed, they are surprising—at least, they were to the authors. But it is not necessary to have a robust intuition for

**Profunctor Optics**

how they work. The important points are first, that *lensC2P* and *lensP2C* are inverses, so the two representations are equivalent (see Appendix C for the proofs); and second, that the profunctor representation supports composition of optics, which we will see in Section 5.

### 4.3 Profunctor prisms

Recall the concrete representation of prisms from Section 1:

**data** *Prism a b s t* = *Prism* {*match* :: $s \to t + a$, *build* :: $b \to t$}

Dually to lenses, the occurrence of the sum type for *match* suggests that the analogue of *Prism* will have something to do with co-cartesian profunctors. Indeed, we define:

**type** *PrismP a b s t* = $\forall p$ . *Cocartesian p* $\Rightarrow$ *Optic p a b s t*

That is, a profunctor prism *PrismP A B S T* lifts a transformer *P A B* on components to a transformer *P S T* on structures, for arbitrary co-cartesian profunctor *P*.

Concrete prisms are themselves co-cartesian profunctors:

**instance** *Profunctor* (*Prism a b*) **where**
   *dimap f g* (*Prism m b*) = *Prism* (*plus g id* · *m* · *f*) (*g* · *b*)

**instance** *Cocartesian* (*Prism a b*) **where**
   *left* (*Prism m b*)   = *Prism* (*either* (*plus Left id* · *m*) (*Left* · *Right*)) (*Left* · *b*)
   *right* (*Prism m b*) = *Prism* (*either* (*Left* · *Left*) (*plus Right id* · *m*)) (*Right* · *b*)

Again dually to lenses, the translations back and forth make crucial use of the lifting to sums. For the translation from the concrete to the profunctor representation, given the two functions *match* :: $S \to T + A$ and *build* :: $B \to T$ that constitute the concrete prism, and a transformer $h :: P\ A\ B$ for some co-cartesian profunctor *P*, we have to construct another transformer of type *P S T*. Now, *right h* has type $P\ (C + A)\ (C + B)$ for any type *C*, and in particular for $C = T$. Then it suffices to wrap this transformer in a preprocessor $S \to T + A$ and a postprocessor $T + B \to T$, both of which are easy to construct:

   *prismC2P* :: *Prism a b s t* $\to$ *PrismP a b s t*
   *prismC2P* (*Prism m b*) = *dimap m* (*either id b*) · *right*

For the translation in the opposite direction, we use the lifting approach again. We are given the profunctor lens $l :: Optic\ P\ A\ B\ S\ T$, which will work for an arbitrary co-cartesian profunctor *P*. We note that *Prism A B* is itself a co-cartesian profunctor, so *l* is applicable at the type $P = Prism\ A\ B$. We therefore construct the trivial concrete prism *Prism Right id* :: *Prism A B A B*, and lift it using *l* to a *Prism A B S T* as required:

   *prismP2C* :: *PrismP a b s t* $\to$ *Prism a b s t*
   *prismP2C l* = *l* (*Prism Right id*)

Again, the proof of the pudding is in the facts that these two translations are inverses, so that the representations are equivalent (Appendix C), and that the profunctor representation supports composition (Section 5).

### 4.4 Profunctor traversals

Recall the concrete representation of traversals from Section 2.3:

   **data** *Traversal a b s t* = *Traversal* {*extract* :: *s* → *FunList a b t*}

The key step in the profunctor representation of traversals is to identify a function *traverse* that lifts a transformation *k* :: *P A B* from *A*s to *B*s to act on each of the elements of a *FunList* in order:

   *traverse* :: (*Cocartesian p*, *Monoidal p*) ⇒ *p a b* → *p* (*FunList a c t*) (*FunList b c t*)
   *traverse k* = *dimap out inn* (*right* (*par k* (*traverse k*)))

Informally, *traverse k* uses *out* to analyse the *FunList*, determining whether it is *Done* or consists of *More* applied to a head and a tail; in the latter case (the combinator *right* lifts a transformer to act on the right-hand component in a sum type), it applies *k* to the head and recursively calls *traverse k* on the tail; then it reassembles the results using *inn*. For this inductive definition to be well founded, it is necessary that the *FunList* is finite, and therefore that the structures being traversed are finite too; this is no additional limitation, because data structures supporting a well-behaved traversal are necessarily finite anyway [3].

Traversals may then be represented as optics, in precisely the same form as lenses and prisms only with a stronger type class constraint:

   **type** *TraversalP a b s t* = ∀*p* . (*Cartesian p*, *Cocartesian p*, *Monoidal p*) ⇒ *Optic p a b s t*

This definition makes *TraversalP A B S T* isomorphic to our earlier more direct notion *Traversal A B S T* of traversals. For the translation from *Traversal* to *TraversalP*, we are given *Traversal h* :: *Traversal A B S T* and a transformer *k* :: *P A B* on elements, for some cartesian, co-cartesian, monoidal profunctor *P*, which we have to lift to a transformer *P S T* on containers. We use *traverse* to lift *k* to obtain a transformer on *FunList*s, which we then sandwich between preprocessor *h* :: *S* → *FunList A B T* and postprocessor *fuse* :: *FunList B B T* → *T* to obtain a transformer on actual containers:

   *traversalC2P* :: *Traversal a b s t* → *TraversalP a b s t*
   *traversalC2P* (*Traversal h*) *k* = *dimap h fuse* (*traverse k*)

In the opposite direction, we have an optic *l* :: *P A B* → *P S T*, applicable for an arbitrary cartesian, co-cartesian, monoidal profunctor *P*, and an effectful operation *m* :: *A* → *F B* on elements for some applicative functor *F*, which we have to lift to a traversal *S* → *F T* over the whole container. Fortunately, *Traversal A B* is itself a cartesian, co-cartesian, monoidal profunctor:

   **instance** *Profunctor* (*Traversal a b*) **where**
     *dimap f g* (*Traversal h*) = *Traversal* (*fmap g* · *h* · *f*)
   **instance** *Cartesian* (*Traversal a b*) **where**
     *first* (*Traversal h*)   = *Traversal* (λ(*s*, *c*) → *fmap* (, *c*) (*h s*))

**Profunctor Optics**

```
    second (Traversal h) = Traversal (λ(c,s) → fmap (c, ) (h s))
instance Cocartesian (Traversal a b) where
    left (Traversal h)  = Traversal (either (fmap Left · h) (Done · Right))
    right (Traversal h) = Traversal (either (Done · Left) (fmap Right · h))
instance Monoidal (Traversal a b) where
    par (Traversal h) (Traversal k) = Traversal (pair h k)
    empty = Traversal pure
```

We can therefore instantiate $P$ to *Traversal A B*; it suffices to take *single*::$A → FunList\ A\ B\ B$, for which *Traversal single* is a trivial concrete traversal *Traversal A B A B*, and use $l$ to lift this to *Traversal A B S T* as required.

```
traversalP2C :: TraversalP a b s t → Traversal a b s t
traversalP2C l = l (Traversal single)
```

These two translations are each other's inverses, as shown in Appendix C. (We need the *Cartesian* constraint for the proofs of equivalence, if not for these definitions.)

In order to apply a traversal, it is useful to define an additional combinator *traverseOf* that turns a profunctor traversal into the kind of traversing function originally described in Section 2.2:

```
traverseOf :: TraversalP a b s t → (∀f . Applicative f ⇒ (a → f b) → s → f t)
traverseOf p f = unUpStar (p (UpStar f))
```

## 5 Composing profunctor optics

Let us return now to the motivating examples from Section 1, where we saw that concrete representations of optics do not support composition well. Things are, happily, much better with the profunctor representation. For example, recall the concrete representation $\pi_1$::*Lens a b* $(a \times c)\ (b \times c)$ of the lens onto the first component of a pair. It is straightforward to translate this lens into the profunctor representation, with $\pi P_1 = lensC2P\ \pi_1$::*LensP a b* $(a \times c)\ (b \times c)$. It is instructive to simplify this definition to first principles; expanding the definitions, we conclude that

```
πP₁ :: Cartesian p ⇒ p a b → p (a × c) (b × c)
πP₁ = dimap (fork fst id) (cross id snd) · first
```

The point here is that the definition of the profunctor lens is not complicated; however, neither is it obvious. Crucially, though, this definition supports composition trivially: since profunctor lenses are nothing but functions, they compose using function composition. In particular, the lens onto the left-most component of a nested pair, which presented us with difficulties in Section 1, may be written directly in terms of $\pi P_1$:

```
πP₁₁ :: LensP a b ((a × c) × d) ((b × c) × d)
πP₁₁ = πP₁ · πP₁
```

Similarly, recall the prism *the* :: *Prism a b* (*Maybe a*) (*Maybe b*) onto an optional value. Its concrete representation can again be directly translated to the profunctor representation:

 *theP* :: *PrismP a b* (*Maybe a*) (*Maybe b*)
 *theP* = *prismC2P the*

or, unpacking the definitions,

 *theP* = *dimap* (*maybe* (*Left Nothing*) *Right*) (*either id Just*) · *right*

(where *maybe* :: $b \to (a \to b) \to Maybe\ a \to b$ deconstructs an optional value). And again, it is a profunctor optic, so it is nothing but a function, and may be combined with other optics using familiar function composition. For example, we may obtain an optic onto the first component of an optional pair:

 *theP* · $\pi P_1$ :: (*Cartesian p*, *Cocartesian p*) $\Rightarrow$ *Optic p a b* (*Maybe* ($a \times c$)) (*Maybe* ($b \times c$))

Note that in a sense the optic is constructed inside out—$\pi P_1$ gives access to a component inside a pair, and *theP* gives access to this pair inside a *Maybe*—whereas the more natural naming is arguably of the form 'first projection of the optional value'. By composing the optics in the opposite order, we obtain instead a composite optic onto the optional first component of a pair:

 $\pi P_1$ · *theP* :: (*Cartesian p*, *Cocartesian p*) $\Rightarrow$ *Optic p a b* (*Maybe a* $\times c$) (*Maybe b* $\times c$)

In both cases, we get the conjunction of the *Cartesian* constraint of lenses and the *Cocartesian* constraint of prisms. Neither combination is purely a lens or purely a prism; they are not expressible using the concrete representations.

We can act on the component in focus under the optic. Specifically, the function arrow ($\to$) is a profunctor, and a *Cartesian*, *Cocartesian*, and *Monoidal* one to boot; so any optic may be applied to a plain function, and will modify the components in focus using that function. For example, to square the integer in the left-hand component of an optional pair, we have

 (*theP* · $\pi P_1$) *square* (*Just* (3, *True*)) = *Just* (9, *True*)

Traversals fit neatly into the scheme too. Recall from Section 2.2 the concrete representation *inorderC* :: *Traversal a b* (*Tree a*) (*Tree b*) of the in-order traversal of an internally labelled binary tree. This is straightforwardly translated into the profunctor representation:

 *inorderP* :: *TraversalP a b* (*Tree a*) (*Tree b*)
 *inorderP* = *traversalC2P inorderC*

and may then be composed with other profunctor optics, using ordinary function composition. Thus, if the tree is labelled with pairs, and we want to traverse only the first components of the pairs, we can use:

**Profunctor Optics**

$inorderP \cdot \pi P_1 :: TraversalP\ a\ b\ (Tree\ (a \times c))\ (Tree\ (b \times c))$

Once we have given such an optic specifying how to access the elements in the tree, we can actually 'apply' the optic by turning it back into a traversal function using *traverseOf*. For example, applying it to the body function *countOdd* will yield a traversal of trees of pairs whose first components are integers, counting the odd ones and returning their parities:

$traverseOf\ (inorderP \cdot \pi P_1)\ countOdd :: Tree\ (Integer \times c) \rightarrow State\ Integer\ (Tree\ (Bool \times c))$

Similarly, we can compose *inorderP* with a prism, *the*, to count the odd integers in a tree which optionally contains values at its nodes:

$traverseOf\ (inorderP \cdot the)\ countOdd :: Tree\ (Maybe\ a) \rightarrow State\ Integer\ (Tree\ (Maybe\ a))$

These examples highlight the common pattern of programming with optics. We compositionally describe which parts of the data we want to access, and separately specify the operation we want to perform. The type system ensures that we can construct only appropriate combinations.

As a slightly more extended example, consider a *Book* of contact details, consisting of a *Tree* of *Entry*s, where each *Entry* has a *Name* and *Contact* details, the latter being either a *Phone* number or a *Skype* identifier:

```
type Number  = String
type ID      = String
type Name    = String
data Contact = Phone Number | Skype ID
data Entry   = Entry Name Contact
type Book    = Tree Entry
```

It is straightforward to define a prism to access the possible phone number in a *Contact*, and a lens to access the *Contact* in an *Entry*—for example, by defining a concrete prism and lens, respectively, and translating each to the profunctor representation:

```
phone :: PrismP Number Number Contact Contact
phone = prismC2P (Prism m Phone) where
   m (Phone n) = Right n
   m (Skype s) = Left (Skype s)
contact :: LensP Contact Contact Entry Entry
contact = lensC2P (Lens v u) where
   v (Entry n c)     = c
   u (c', Entry n c) = Entry n c'
```

These may be combined, in order to access the possible phone number in an *Entry*:

```
contactPhone :: TraversalP Number Number Entry Entry
contactPhone = contact · phone
```

and combined further with in-order traversal, in order to access all the phone numbers in a *Book* of contacts:

  *bookPhones* :: *TraversalP Number Number Book Book*
  *bookPhones* = *inorderP* · *contact* · *phone*

If we have a function *tidyNumber* :: *Number* → *Number* to normalize a phone number, perhaps to add spaces, parentheses, and hyphens according to local custom, then we can tidy a whole *Book* by tidying each phone number:

  *tidyBook* :: *Book* → *Book*
  *tidyBook* = *bookPhones tidyNumber*

If we have a function *output* :: *Number* → *IO Number* in the *IO* monad of input–output actions, which prints out a phone number and returns a copy of it too, then we can print each of the numbers in turn:

  *printBook* :: *Book* → *IO Book*
  *printBook* = *traverseOf bookPhones output*

There is an applicative functor *Const m* for any monoid *m*, and in particular for the list monoid; using this, we can extract a list of all the phone numbers in a book of contacts:

  *listBookPhones* :: *Book* → [*Number*]
  *listBookPhones* = *getConst* · *traverseOf bookPhones* (*Const* · (λ$x$ → [$x$]))

## 6  Related work

Lenses were introduced by Foster, Pierce *et al.* [9] as a model of bidirectional transformations. Their motivation was to take a linguistic approach to the so-called *view–update problem* in databases [2], the problem there being, given a computed view table (analogous to the *view* method of our *Lens*), to propagate a modified view back as a corresponding update to the original source tables (analogous to our *update*).

A basic criterion for soundness of a bidirectional transformation is that it satisfies a pair of round-trip laws. In our terminology, this criterion presupposes a monomorphic lens *Lens A A S S*, in which the types do not change. Given *Lens v u* of this type, the first law is that $v\,(u\,(a,s)) = a$; informally, that a modified view is faithfully stored, so that it may later be retrieved. The second is that $u\,(v\,s,s) = s$; informally, that propagating an unmodified view does not change the source. Such a lens is said to be *well behaved*. (As we discuss in Section 7, well-behavedness is orthogonal to the question of whether or not the profunctor representation matches the concrete one.)

In the simple case, the source factorizes cleanly into the view and its 'complement', for example when the view is the left component of a pair; in such cases, more can be said. Specifically, *Lens v u* will satisfy a third law, that $u\,(a',(u\,(a,s))) = u\,(a',s)$;

informally, a later update completely overwrites an earlier one. In that case, the lens is said to be *very well behaved*; in the database community, this is called a *constant complement* situation, because the complement is untouched by an update. This special case received earlier attention from programming language researchers, for example in the work of Kagawa [15] on representing mutable cells in pure languages, and of Oles [29] on modelling block structure in Algol-like languages.

One can consider programming with optics as an application of *datatype-generic programming* [11], that is, the construction of programs that are parametrized by the shape of the data they manipulate. The parameter has traditionally been a functor; for us, it is a profunctor. In particular, there is significant related work on datatype-generic traversals. The essential structure of the generic traversal is already apparent in the work of Meertens [24]. Other approaches, including Lämmel's *Scrap Your Boilerplate* (SYB) [21], Mitchell's uniplate library [25], Bringert's Compos library [5], McBride and Paterson's *Traversable* class [23], and O'Connor's Multiplate [26], all provide traversal functions of similar forms to our definition of *traverse*. In recent years, attention has turned to the in-depth comparison and study of these different definitions [3, 12, 14].

The SYB approach [21] is of particular note due to the recognition of the need to combine an effectful traversal with an operation that focusses in turn on each element of a data structure. The SYB implemention of this is ad-hoc, using dynamic type checking, but in our framework we can write the same programs by composing a traversal with a lens. Thus, another way to view our work is as a generalisation of effectful traversals.

More recent work has explored the so-called van Laarhoven representation [20, 26] in terms of functions of type $(A \to F\,B) \to (S \to F\,T)$ for various functors $F$, which is the predominent representation currently used in Haskell. This representation shares many properties with the profunctor representation we describe, but is slightly less elegant (it requires instantiation of the functor $F$ even when it is not needed—for example, for simple adapters). Other representations of lenses have also been explored [22], but these appear to lack extension to other varieties of optic at all.

As far as we are aware, prisms have not previously been described in the literature, and are only folklore amongst functional programmers. Reid Barton is credited by Edward Kmett [16] with the observation that there should exist a 'dual' to lenses which works with sum types. The development of the idea was then led by Kmett, Elliott Hird and Shachaf Ben-Kiki whilst working on the lens library [17]. Prisms are an implementation of first-class patterns, unlike other proposals; for example, Tullsen [37] recognised the close connection between data constructors and pattern matching. Pattern matching is the method which is used to deconstruct values constructed by data constructors; first-class patterns should also be able to re-build the values that they match. Pattern synonyms [32] are also bidirectional in this sense, but are not first-class. Scala has the closely related *extractor objects* [8] which have two special methods *unapply* and *apply* which correspond to matching and building respectively.

What is unique to the framework we have described is the explicit connection between these different kinds of generic functions. This is further highlighted by the representation allowing us to seamlessly upcast and use more specific optics in places where a less powerful optic would suffice—for example, using a lens as a traversal.

Bringing together these different styles of datatype-generic programming makes it straightforward to construct heterogeneous composite data accessors, a use case that is not possible in each framework individually.

There are several nascent implementations of profunctor-based optics. The most well developed is the purescript library purescript-profunctor-optics [10], which provides indexed optics in addition to the optics that we have described. The optics library [7] is a Javascript proof of concept implementation. Russell O'Connor's Haskell implementation mezzolens [27] was instrumental in our understanding.

## 7  Discussion

We have drawn together a series of folklore developments that together lead to a modular framework for data accessors. This framework accommodates *adapters*, which provide access via a change of representation; *lenses*, which provide access to a component of a product structure, such as a field of a record; *prisms*, which provide access to a component of a sum structure, such as one variant from a union; and *traversals*, which provide access to a sequence of components, such as the elements in a container datatype. Collectively, these four varieties of data accessor are called *optics*. Crucially, the four varieties of optic have a similar representation, and this form is closed under composition; this allows us to combine different varieties of optic, such as a lens with a prism, which is not possible with more direct representations.

The particular representation we use is *mappings between transformers*, where transformers are represented in terms of *profunctors*, a generalization of functions:

$$\forall p \,.\, \mathit{Profunctor}\ p \Rightarrow p\ a\ b \to p\ s\ t$$

In other words, it is a representation using *higher-order functions* rather than more concrete datatypes. This choice of representation is the essential trick, both accommodating the wide variety of apparently distinct optics, and straightforwardly supporting combinations via function composition. That this representation is even adequate comes as quite a surprise—it is salutary to reflect on Christopher Strachey's observation of half a century ago [34] that

> *many of the more interesting developments of programming and programming languages come from the unrestricted use of functions, and in particular of functions which have functions as results* and yet we are still finding new applications of higher-order functions.

It is interesting to note that the four representations we have chosen for the four different varieties of optic form a lattice, as shown in Figure 5: adapters are special kinds of lens and of prism, and lenses and prisms are each special kinds of traversal. They all have the higher-order functional form quoted above, differing only in the constraints imposed on the parameter *p*. A combination of different varieties of optic is also of the same form, but collects all the constraints from the individual parts; that is, it forms an upper bound in the lattice. Thus, the combination of an adapter with a lens is another lens, and the combination of a traversal with anything is again a traversal.

**Profunctor Optics**

The lattice structure becomes apparent only in the profunctor representation, because heterogeneous combinations are not otherwise expressible. The combination of a lens and a prism is a traversal; but in fact, the combination needs only the *Cartesian* and *Cocartesian* constraints of lenses and prisms respectively, and not the additional *Monoidal* constraint of traversals, so does not use the full power of a traversal (indeed, such a combination necessarily targets at most one component of a structure, and so there is no need for sequencing of effectful operations). This means that there is a fifth point in the lattice, the least upper bound of lenses and prisms but strictly below traversals. It is as yet unclear to us whether that fifth point is a useful abstraction in its own right, or a mere artifact of our representation; this question calls for further work.

Curiously, our presentation does not depend at all on the 'well-behavedness' laws of lenses or their duals for prisms, nor on the two functions making up an adapter being each other's inverses, nor on the laws of traversals [12]. The proofs of equivalence (in Appendix C) do not use them; the abstractions accommodate ill-behaved optics just as well as they do well-behaved ones. Identifying suitable well-behavedness laws for profunctor optics is another topic for future research. We conjecture that addressing this question will entail consideration of sequential composition of profunctors (taking a $P\ A\ B$ and a $P\ B\ C$ to a $P\ A\ C$). That might have other benefits too; in particular, given composition, one can define *par* in terms of *first* and *second*, which should simplify the assumptions we make.

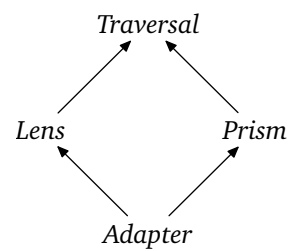

**Figure 5** The lattice of profunctor optics (arrow $X \to Y$ denotes that $X$ is a subclass of $Y$).

We have taken the opportunity to judiciously rename some abstractions from existing libraries, in the interests of tidying up. The names *first*, *second*, *left*, and *right* were already popular in the Haskell *Arrow* library, so Kmett's *Profunctor* library [18] uses *first'*, *second'*, *left'*, *right'* instead. What we have called an *Adapter* is conventionally called an *Iso*, despite there being no requirement for it actually to form an isomorphism. Our classes *Cartesian* and *Cocartesian* are conventionally (and asymmetrically) called *Strong* and *Choice*. The conventional ordering of type parameters for optics, for example from the mezzolens library [27], would be to have *Lens s t a b* and so on; we have used the ordering *Lens a b s t* instead, so that we can conveniently apply the type constructor just to the first two, as required for the profunctor instances in the translation functions in Section 4. Our *Traversal* is not quite the same as the *traverse* method of the *Traversable* type class in the Haskell libraries, because that type class insists that the container datatype is polymorphic; ours allows us for example to access a packed *ByteString* as a container of eight-bit *Word8*s, or an integer as a container of digits—traversal still makes sense for monomorphic containers, it just cannot change the types of the elements.

Although we have presented our constructions using Haskell as a vehicle, they do not really depend in any essential way on Haskell. All that seems to be required are higher-order functions ('lambdas' or 'closures'), parametrized types ('generics' or 'abstract types') of higher kind (in particular, parametrization by profunctors), and some mechanism for separating interfaces from implementations ('abstract classes' or 'modules'). We feel that Haskell allows clear expression using those features—the *algorithmic language* de nos jours—but it is possible to replicate our constructions in other languages such as Scala; we sketch an implementation in Appendix B.

**Acknowledgements**   Our primary thanks are due to the various people who have contributed to the library development and online discussions whose outcomes we have made use of in this paper, especially Edward Kmett, Russell O'Connor, Twan van Laarhoven, Reid Barton, Shachaf Ben-Kiki, and Elliott Hird. We are grateful to Paul Levy for suggesting the names 'cartesianly strong', 'co-cartesianly strong', and 'monoidal', and to Sam Staton, Guillaume Boisseau, and especially James McKinna for many helpful comments and encouraging discussions. This paper is a condensed version of the first author's undergraduate thesis [31]. The work has been partially supported by the UK EPSRC-funded project *A Theory of Least Change for Bidirectional Transformations* (EP/K020919/1).

**Profunctor Optics**

## A  Haskell background

We have used an idealization of Haskell [30] throughout as a lingua franca, although our constructions do not really depend on anything other than higher-order functions and some notion of interface and implementation. For the reader unfamiliar with Haskell, we summarize here some conventions and useful standard operations. We also explain our idealizations, so even those familiar with Haskell should skim this section.

Haskell is a functional programming language, so of course revolves around functions. The type $A \rightarrow B$ consists of functions that take an argument of type $A$ and return a result of type $B$; for example, the function *even* of type $Integer \rightarrow Bool$ determines whether or not an *Integer* is even. Type declarations are written with a double colon:

```
even :: Integer → Bool
even n = (mod n 2 == 0)
```

The identity function is written *id*, and function composition with a centred dot, so that $(f \cdot g)\,x = f\,(g\,x)$.

By convention, functions are *curried* wherever possible; for example, rather than defining the modulus function to accept a pair of arguments:

```
mod :: (Integer, Integer) → Integer
```

we make it take two separate arguments:

```
mod :: Integer → Integer → Integer
```

or equivalently, to take one argument and yield a function of the other:

```
mod :: Integer → (Integer → Integer)
```

Functions may be *polymorphic*; so identity and composition have the following types:

```
id  :: a → a
(·) :: (b → c) → (a → b) → (a → c)
```

Type variables (such as *a* above) are written with a lowercase initial letter, and are implicitly universally quantified; specific types (such as *Integer*) are written with an uppercase initial letter. We therefore use uppercase identifiers (such as $A, B, C$) for specific types in examples in prose.

Pairs are written in parentheses, both as values and as types; for example, $(3, True)::$ $(Integer, Bool)$. However, we have used × for pair types in the paper, writing $Integer \times Bool$ for $(Integer, Bool)$. Two functions with a common source type may be combined to yield a pair of results, via the function *fork*:

```
fork :: (a → b) → (a → c) → a → (b, c)
fork f g x = (f x, g x)
```

**Profunctor Optics**

and two functions may act in parallel on the components of a pair, via the function *cross*:

$$\mathit{cross} :: (a \to a') \to (b \to b') \to (a,b) \to (a',b')$$
$$\mathit{cross}\ f\ g\ (x,y) = (f\ x, g\ y)$$

Since pairs have projections $\mathit{fst} :: (a,b) \to a$ and $\mathit{snd} :: (a,b) \to b$, the *cross* operation is in fact an instance of *fork*:

$$\mathit{cross}\ f\ g = \mathit{fork}\ (f \cdot \mathit{fst})\ (g \cdot \mathit{snd})$$

Dually, *sum* types correspond to 'variant' types in Pascal or unions in C. In Haskell, they are written with an *algebraic datatype*:

**data** *Either a b* = *Left a* | *Right b*

(In the paper, we have used the idealized notation $A + B$ for *Either A B*.) This declaration introduces a new two-parameter (polymorphic) datatype *Either*, and two (polymorphic) constructors $\mathit{Left} :: a \to \mathit{Either}\ a\ b$ and $\mathit{Right} :: b \to \mathit{Either}\ a\ b$; so values of type *Either A B* are either of the form *Left x* with $x :: A$, or of the form *Right y* with $y :: B$. Two functions with a common target type may be combined to act on a sum type, via the function *either*:

$$\mathit{either} :: (a \to c) \to (b \to c) \to \mathit{Either}\ a\ b \to c$$
$$\mathit{either}\ f\ g\ (\mathit{Left}\ x)\ \ = f\ x$$
$$\mathit{either}\ f\ g\ (\mathit{Right}\ y) = g\ y$$

As an instance of this, we can act independently on the two variants of a sum, via the function *plus*:

$$\mathit{plus} :: (a \to a') \to (b \to b') \to \mathit{Either}\ a\ b \to \mathit{Either}\ a'\ b'$$
$$\mathit{plus}\ f\ g = \mathit{either}\ (\mathit{Left} \cdot f)\ (\mathit{Right} \cdot g)$$

A useful special case of sum types is sum with the unit type, which provides a representation of *optional* values:

**data** *Maybe a* = *Just a* | *Nothing*

It is sometimes convenient to model a datatype as a *record* with named fields; for example, had pairs not been built in, we might have defined instead

**data** *Pair a b* = *MkPair* {*fst* :: *a*, *snd* :: *b*}

This declaration introduces a new two-parameter datatype *Pair*, a constructor $\mathit{MkPair} :: a \to b \to \mathit{Pair}\ a\ b$, and two field extractors $\mathit{fst} :: \mathit{Pair}\ a\ b \to a$ and $\mathit{snd} :: \mathit{Pair}\ a\ b \to b$. The namespaces for types (such as *Pair*) and values (such as *MkPair*) are disjoint, and it is idiomatic Haskell to name the constructor *Pair* rather than *MkPair*.

Algebraic datatypes find most use for recursive datatypes. For example, a polymorphic datatype of internally labelled binary trees may be defined by

7-32

**data** *Tree a* = *Empty* | *Node* (*Tree a*) *a* (*Tree a*)

which has a constant *Empty* representing the empty tree, and a ternary constructor *Node* assembling a non-empty tree from two children and a root label. Haskell provides a built-in polymorphic datatype [*a*] of lists; but if that had not been provided, we could have defined instead

**data** *List a* = *Nil* | *Cons a* (*List a*)

*Interfaces and implementations* are typically modelled in Haskell by means of *type classes*. A type class represents a set of types, characterized by their common support for a particular collection of *methods*. For example, the standard type classes *Eq* and *Ord* denote the classes of types that support equality and ordering functions:

**class** *Eq a* **where**
  (==) :: $a \to a \to Bool$
**class** *Eq a* $\Rightarrow$ *Ord a* **where**
  ($\leqslant$)  :: $a \to a \to Bool$

This declaration states that a type *A* is a member of the type class *Eq* if it supports an equality test (==) :: $A \to A \to Bool$; and an *Eq* type *A* is a fortiori an *Ord* type if it also supports the comparison ($\leqslant$) :: $A \to A \to Bool$. Most of the standard types are instances of these classes. New instances may be declared by providing a definition of the necessary methods; for example, we could specify equality on optional values, given equality on their elements, by:

**instance** *Eq a* $\Rightarrow$ *Eq* (*Maybe a*) **where**
  *Just x*    == *Just y*   = (*x* == *y*)
  *Nothing* == *Nothing* = *True*
  _          == _         = *False*

(where the underscore is a wild-card pattern). To be precise, one typically declares laws that instance methods should satisfy (such as being an equivalence relation for *Eq*, and a preorder for *Ord*); but Haskell provides no way to state such laws other than in comments. One can think of the type class as representing an interface, and the instances as implementations of that interface.

Type classes may be used for *type constructors* as well as for concrete types. For example, we make extensive use of the type class *Functor*, which represents polymorphic *container types*. Informally, the type constructor *F* represents a polymorphic container if a value of type *F A* contains elements of type *A*, on which one may operate with a function of type $A \to B$, to yield a new value of type *F B*. Formally:

**class** *Functor f* **where**
  *fmap* :: $(a \to b) \to f\ a \to f\ b$

For example, the *Maybe* type constructor is a functor:

**instance** *Functor Maybe* **where**
  *fmap f* (*Just x*)  = *Just* (*f x*)
  *fmap f Nothing* = *Nothing*

which should satisfy the following two laws:

$$fmap\ (f \cdot g) = fmap\ f \cdot fmap\ g$$
$$fmap\ id\quad = id$$

An important result about polymorphic functions between functors is known colloquially as *theorems for free* [40]. This result entails that a function $h$ of type $\forall a\,.\,F\ a \to G\ a$, for particular $F$ and $G$ that are instances of *Functor*, but crucially for *all* values of the type parameter $a$, satisfies the property

$$h \cdot fmap_F\ f = fmap_G\ f \cdot h$$

where the occurrences of *fmap* have been annotated to indicate that on the left of the equation it is the instance for $F$ and on the right it is the one for $G$, as may be easily be determined from the type of $h$. We call this property 'the free theorem of the type of $h$', or sometimes just 'the free theorem of $h$' when we have a particular definition of $h$ in mind. The point is that it does not matter what that definition is, provided that it has the stated type. For example, consider a function $h :: [a] \to Maybe\ a$. It does not matter whether this returns the head of the list and *Nothing* for the empty list, or the last element of the list similarly, or the fourth element when the list has odd length at least 13 and *Nothing* for even-length and shorter lists; it still necessarily satisfies the same property. (To be precise, the theorem may need a side condition when $h$ is allowed to be a partial function; but in this paper, we restrict attention to total functions.)

Free theorems also apply to types involving constraints [38], essentially by following the translation from type classes to dictionary-passing style—the functions being mapped must preserve the methods of the class. For example, the function $nub :: Eq\ a \Rightarrow [a] \to [a]$ removes duplicates from a list. The free theorem for a function $h$ of type $[a] \to [a]$ would state that $h \cdot fmap_{[\,]}\ f = fmap_{[\,]}\ f \cdot h$ for any $f$, which is plainly false for $h = nub$. Of course, *nub* has a more constrained than $h$; the expected theorem restricts the claim to those $f$ that preserve equality—that is, such that $(f\ x == f\ y) \iff (x == y)$. This is in fact precisely the theorem that arises for free from the type of the function $nubBy :: (a \to a \to Bool) \to [a] \to [a]$, which is essentially the dictionary-passing translation of *nub*.

## B   Alternative implementation

In this appendix, we sketch a parallel implementation in Scala of the concepts introduced in the paper. This exercise demonstrates that the ideas are not limited to Haskell. However, the constructions do require support for higher-kinded types (specifically, for instances of the *Profunctor* class), which is not available in languages such as Java and C#.

We require Scala's support for higher-kinded types:

**import** *scala.language.higherKinds*

Abstractions that are represented with type classes in Haskell are traditionally implemented using *traits*—a variant of classes allowing flexible mixin composition—in Scala. The following definition introduces the *Profunctor* abstraction, here a two-parameter type operation $p$. For some $p$ to instantiate the *Profunctor* abstraction, it must implement the *dimap* method. The type of *dimap* is parametrized by four type variables $a, b, c, d$. The method itself takes two functions $f$ and $g$ of types $c \Rightarrow a$ and $b \Rightarrow d$ respectively, and an input transformation $h : p[a, b]$, which it lifts to an output transformation of type $p[c, d]$.

> **trait** *Profunctor*[$p[\_,\_]$]{  
>   **def** *dimap*[$a, b, c, d$]($f : c \Rightarrow a$)($g : b \Rightarrow d$)($h : p[a, b]$) : $p[c, d]$  
> }

Similarly, the *Cartesian*, *Cocartesian*, and *Monoidal* specializations each involve implementing one or two methods. The Scala libraries provide a generic class *Tuple2* of pairs (in fact, the language also supports the Haskell syntax $(a, b)$ as a shorthand for *Tuple2*[$a, b$]), a generic class *Either* precisely matching Haskell's *Either*, and a class *Unit* with a single inhabitant corresponding to Haskell's ().

> **trait** *Cartesian*[$p[\_,\_]$] **extends** *Profunctor*[$p$]{  
>   **def** *first*[$a, b, c$]($h : p[a, b]$) : $p[Tuple2[a, c], Tuple2[b, c]]$  
> }  
> **trait** *Cocartesian*[$p[\_,\_]$] **extends** *Profunctor*[$p$]{  
>   **def** *right*[$a, b, c$]($h : p[a, b]$) : $p[Either[c, a], Either[c, b]]$  
> }  
> **trait** *Monoidal*[$p[\_,\_]$] **extends** *Profunctor*[$p$]{  
>   **def** *par*[$a, b, c, d$]($h : p[a, b]$)($k : p[c, d]$) : $p[Tuple2[a, c], Tuple2[b, d]]$  
>   **def** *empty* : $p[Unit, Unit]$  
> }

The four varieties of optic are each represented as abstract classes, each with a single method called *apply* as the operative content. What is represented with a type class constraint in Haskell turns up as an *implicit* parameter on the *apply* methods in Scala; when there is a unique binding of the appropriate type in scope at the call site, the actual parameter may be omitted and this binding will be used in its place. In the case of *Traversal*, the implicit parameter is required to implement the mixin composition of three traits, analogous to having three type class constraints in Haskell.

> **abstract class** *Adapter*[$p[\_,\_], a, b, s, t$]{  
>   **def** *apply*($h : p[a, b]$)(**implicit** $prof : Profunctor[p]$) : $p[s, t]$  
> }  
> **abstract class** *Lens*[$p[\_,\_], a, b, s, t$]{  
>   **def** *apply*($h : p[a, b]$)(**implicit** $prof : Cartesian[p]$) : $p[s, t]$  
> }  
> **abstract class** *Prism*[$p[\_,\_], a, b, s, t$]{  
>   **def** *apply*($h : p[a, b]$)(**implicit** $prof : Cocartesian[p]$) : $p[s, t]$  

**Profunctor Optics**

}

**abstract class** *Traversal*[$p$[_, _], $a, b, s, t$]{
  **def** *apply*($h : p[a, b]$)
        (**implicit** *prof* : *Cartesian*[$p$] **with** *Cocartesian*[$p$] **with** *Monoidal*[$p$]) : $p[s, t]$
}

Here is a concrete class corresponding to the lens $\pi_1$ we used in the paper, for access to the left-hand component of a pair. It is defined by extending a suitable type instantiation of the *Lens* trait, and providing an implementation of the *apply* method. The three subsidiary definitions *view*, *update*, and *fork* are local to *apply*; the final line of the definition of *apply* is the actual body. The value is obtained by applying of *dimap* to *first*, both of which are methods of the *Cartesian* profunctor *prof*.

  **class** *PiOne*[$p$[_, _], $a, b, c$] **extends** *Lens*[$p, a, b$, *Tuple2*[$a, c$], *Tuple2*[$b, c$]]{
    **def** *apply*($f : p[a, b]$)(**implicit** *prof* : *Cartesian*[$p$]) : $p$[*Tuple2*[$a, c$], *Tuple2*[$b, c$]] = {
      **def** *view*[$a, c$] : (*Tuple2*[$a, c$] $\Rightarrow a$) =
        ($xy \Rightarrow xy\_1$);
      **def** *update*[$a, b, c$] : (*Tuple2*[$b$, *Tuple2*[$a, c$]] $\Rightarrow$ *Tuple2*[$b, c$]) =
        ($zxy \Rightarrow$ *newTuple2*($zxy\_1, (zxy\_2)\_2$));
      **def** *fork*[$a, b, c$]($f : a \Rightarrow b$)($g : a \Rightarrow c$) : ($a \Rightarrow$ *Tuple2*[$b, c$]) =
        ($x \Rightarrow$ *newTuple2*($f(x), g(x)$));
      *prof*.*dimap*(*fork*(*view*[$a, c$])(*identity*))(*update*[$a, b, c$])(*prof*.*first*($f$))
    }
  }

Similarly, here is a class corresponding to the prism *the* from the body of the paper, for access to the payload of an optional value. The Scala libraries provide a class *Option* corresponding to Haskell's *Maybe* type constructor. The local functions *match* and *either* are defined using case analysis over one argument. (Scala uses **match** as a keyword, so we have taken a liberty in using that word as an identifier.)

  **class** *The*[$p$[_, _], $a, b$] **extends** *Prism*[$p, a, b$, *Option*[$a$], *Option*[$b$]]{
    **def** *apply*($f : p[a, b]$)(**implicit** *prof* : *Cocartesian*[$p$]) : $p$[*Option*[$a$], *Option*[$b$]] = {
      **def** *match*[$a, b$] : (*Option*[$a$] $\Rightarrow$ *Either*[*Option*[$b$], $a$]) = {
        **case** *Some*($a$) $\Rightarrow$ *Right*($a$)
        **case** *None*    $\Rightarrow$ *Left*(*None*)
      };
      **def** *build*[$b$] : ($b \Rightarrow$ *Option*[$b$]) = ($b \Rightarrow$ *Some*($b$));
      **def** *either*[$a, b, c$]($f : a \Rightarrow c$)($g : b \Rightarrow c$) : (*Either*[$a, b$] $\Rightarrow c$) = {
        **case** *Left*($a$)   $\Rightarrow f(a)$
        **case** *Right*($b$) $\Rightarrow g(b)$
      };
      *prof*.*dimap*(*match*[$a, b$])(*either*(*identity*[*Option*[$b$]])(*build*[$b$]))(*prof*.*right*($f$))
    }
  }

As well as Haskell, there are implementations of variations of these ideas in Javascript [1, 7] and Purescript [10]. There is a Scala library cats that implements profunctors but has not been extended to optics [6], and a Scala library monocle of optics that does not use profunctors [35]; there are some initial experiments towards a Scala implementation of profunctor optics [36] that does not yet seem to be complete. There is also a Java library called Functional Java [13] implementing optics without using profunctors, so without their modularity benefits.

## C  Proofs of equivalence

We formalize here the statements made earlier about the equivalences between the concrete and profunctor representations of the various kinds of optic, and provide proofs of those equivalences.

**Theorem 1.** The functions *adapterC2P* and *adapterP2C* are each other's inverses, and so the types *Adapter A B S T* and *AdapterP A B S T* are isomorphic for all type parameters $A, B, S, T$. □

**Theorem 2.** The functions *lensC2P* and *lensP2C* are each other's inverses, and so the types *Lens A B S T* and *LensP A B S T* are isomorphic for all type parameters $A, B, S, T$. □

**Theorem 3.** The functions *prismC2P* and *prismP2C* are each other's inverses, and so the types *Prism A B S T* and *PrismP A B S T* are isomorphic for all type parameters $A, B, S, T$. □

**Theorem 4.** The functions *traversalC2P* and *traversalP2C* are each other's inverses, and so the types *Traversal A B S T* and *TraversalP A B S T* are isomorphic for all type parameters $A, B, S, T$. □

### C.1  Adapters

One of the key ingredients in the proofs of the theorems is the notion of *profunctor morphism*.

**Definition 5.** A polymorphic function $phi :: \forall a\, b\, .\, P\, a\, b \to Q\, a\, b$, for given profunctors $P, Q$ but for all types $a, b$, is a 'profunctor morphism from $P$ to $Q$' if

$$dimap_Q\, f\, g \cdot phi = phi \cdot dimap_P\, f\, g$$

for all functions $f, g$, where we have annotated the two occurrences of *dimap* to indicate which instances they are. □

In particular, the translation function *adapterC2P* from Section 4.1, when applied to a given transformer of type $P\, A\, B$ as its second argument, yields a profunctor morphism from *Adapter A B* to $P$. To make this precise, consider

$$flip\ adapterC2P :: Profunctor\ p \Rightarrow p\ a\ b \to Adapter\ a\ b\ s\ t \to p\ s\ t$$

which is like *adapterC2P* but takes its arguments in a different order, since *flip f x y* = *f y x*.

Then we may state:

**Lemma 6.** For given $k :: P\ A\ B$ for some profunctor $P$ and types $A, B$, the function *flip adapterC2P k* :: *Adapter A B s t* $\to$ *P s t* is a profunctor morphism from *Adapter A B* to $P$. □

**Proof.** We have:

    *dimap f g* (*flip adapterC2P k* (*Adapter o i*))
= 〚 *flip* 〛
    *dimap f g* (*adapterC2P* (*Adapter o i*) *k*)
= 〚 *adapterC2P* 〛
    *dimap f g* (*dimap o i k*)
= 〚 *dimap* composition 〛
    *dimap* (*o* · *f*) (*g* · *i*) *k*
= 〚 *adapterC2P* 〛
    *adapterC2P* (*Adapter* (*o* · *f*) (*g* · *i*)) *k*
= 〚 *dimap* for *Adapter* 〛
    *adapterC2P* (*dimap f g* (*Adapter o i*)) *k*
= 〚 *flip* 〛
    *flip adapterC2P k* (*dimap f g* (*Adapter o i*))

and so *dimap f g* · *flip adapterC2P k* = *flip adapterC2P k* · *dimap f g* as required. ♡

The second key ingredient is the *free theorem* of profunctor optics, obtained by instantiating Wadler's 'theorems for free' at the appropriate type [38, 40]:

**Lemma 7.** For any $l$ of type $\forall p\ .\ Profunctor\ p \Rightarrow Optic\ p\ A\ B\ S\ T$ and any profunctor morphism *phi* from profunctor $P$ to profunctor $Q$, we have $l \cdot phi = phi \cdot l$. □

Both sides are of type $P\ A\ B \to Q\ S\ T$; the free theorem states that lifting the $P\ A\ B$ to $P\ S\ T$ and then translating to $Q\ S\ T$ coincides with first translating to $Q\ A\ B$ and then lifting to $Q\ S\ T$.

**Proof (of Theorem 1).** One direction is quite straightforward:

    *adapterP2C* (*adapterC2P* (*Adapter o i*))
= 〚 *adapterC2P* 〛
    *adapterP2C* (*dimap o i*)
= 〚 *adapterP2C* 〛
    *dimap o i* (*Adapter id id*)
= 〚 *dimap* for *Adapter* 〛
    *Adapter* (*id* · *o*) (*i* · *id*)
= 〚 identity 〛
    *Adapter o i*

as required. For the other direction, we need to use Lemma 6:

    *adapterC2P* (*adapterP2C l*) *k*
= ⟦ *adapterP2C* ⟧
    *adapterC2P* (*l* (*Adapter id id*)) *k*
= ⟦ *flip* ⟧
    *flip adapterC2P k* (*l* (*Adapter id id*))
= ⟦ free theorem of *l*, and Lemma 6 ⟧
    *l* (*flip adapterC2P k* (*Adapter id id*))
= ⟦ *flip* ⟧
    *l* (*adapterC2P* (*Adapter id id*) *k*)
= ⟦ *adapterC2P* ⟧
    *l* (*dimap id id k*)
= ⟦ *dimap* identity ⟧
    *l k*

so *adapterC2P* · *adapterP2C* = *id* as required.     ♡

### C.2 Lenses

**Lemma 8.** The 'free theorem' [40] of *first* is that

*dimap id h k* = *dimap g id l* ⇒ *dimap id* (*cross* (*h*, *f*)) (*first k*) = *dimap* (*cross* (*g*, *f*)) *id* (*first l*)

    □

**Lemma 9.** For given $k :: P\ A\ B$ for some cartesian profunctor $P$ and types $A, B$, the function *flip lensC2P k* :: *Lens A B s t* → *P s t* is a profunctor morphism from *Lens A B* to $P$.     □

**Proof.** We have:

    *dimap f g* (*flip lensC2P k* (*Lens v u*))
= ⟦ *flip* ⟧
    *dimap f g* (*lensC2P* (*Lens v u*) *k*)
= ⟦ *lensC2P* ⟧
    *dimap f g* (*dimap* (*fork* (*v*, *id*)) *u* (*first k*))
= ⟦ *dimap* composition ⟧
    *dimap* (*fork* (*v*, *id*) · *f*) (*g* · *u*) (*first k*)
= ⟦ products and fork ⟧
    *dimap* (*cross* (*id*, *f*) · *fork* (*v* · *f*, *id*)) (*g* · *u*) (*first k*)
= ⟦ *dimap* composition ⟧
    *dimap* (*fork* (*v* · *f*, *id*)) (*g* · *u*) (*dimap* (*cross* (*id*, *f*)) *id* (*first k*))
= ⟦ free theorem of *first* (Lemma 8), with *g* = *id*, *h* = *id*, and *k* = *l* ⟧
    *dimap* (*fork* (*v* · *f*, *id*)) (*g* · *u*) (*dimap id* (*cross* (*id*, *f*)) (*first k*))
= ⟦ *dimap* composition ⟧
    *dimap* (*fork* (*v* · *f*, *id*)) (*g* · *u* · *cross* (*id*, *f*)) (*first k*)

**Profunctor Optics**

$\qquad = \quad [\![ \; \textit{lensC2P} \; ]\!]$
$\qquad \textit{lensC2P} \, (\textit{Lens} \, (v \cdot f) \, (g \cdot u \cdot \textit{cross} \, (\textit{id}, f))) \, k$
$\qquad = \quad [\![ \; \textit{dimap} \text{ for } \textit{Lens} \; ]\!]$
$\qquad \textit{lensC2P} \, (\textit{dimap} \, f \, g \, (\textit{Lens} \, v \, u)) \, k$
$\qquad = \quad [\![ \; \textit{flip} \; ]\!]$
$\qquad \textit{flip} \, \textit{lensC2P} \, k \, (\textit{dimap} \, f \, g \, (\textit{Lens} \, v \, u))$

So $\textit{dimap} \, f \, g \cdot \textit{flip} \, \textit{lensC2P} \, k = \textit{flip} \, \textit{lensC2P} \, k \cdot \textit{dimap} \, f \, g$ as required. ♡

We also need the following observation, due to O'Connor [28].

**Lemma 10.**

$\qquad \textit{dimap} \, (\textit{fork} \, (\textit{id}, \textit{id})) \, \textit{fst} \cdot \textit{first} = \textit{id}$

☐

**Proof.** We have:

$\qquad \textit{dimap} \, (\textit{fork} \, (\textit{id}, \textit{id})) \, \textit{fst} \cdot \textit{first}$
$\quad = \quad [\![ \; \text{products} \; ]\!]$
$\qquad \textit{dimap} \, (\textit{fork} \, (\textit{id}, \textit{id})) \, (\textit{fst} \cdot \textit{cross} \, (\textit{id}, \textit{const} \, ())) \cdot \textit{first}$
$\quad = \quad [\![ \; \textit{dimap} \text{ composition} \; ]\!]$
$\qquad \textit{dimap} \, (\textit{fork} \, (\textit{id}, \textit{id})) \, \textit{fst} \cdot \textit{dimap} \, \textit{id} \, (\textit{cross} \, (\textit{id}, \textit{const} \, ())) \cdot \textit{first}$
$\quad = \quad [\![ \; \text{free theorem of } \textit{first} \; ]\!]$
$\qquad \textit{dimap} \, (\textit{fork} \, (\textit{id}, \textit{id})) \, \textit{fst} \cdot \textit{dimap} \, (\textit{cross} \, (\textit{id}, \textit{const} \, ())) \, \textit{id} \cdot \textit{first}$
$\quad = \quad [\![ \; \textit{dimap} \text{ composition} \; ]\!]$
$\qquad \textit{dimap} \, (\textit{cross} \, (\textit{id}, \textit{const} \, ()) \cdot \textit{fork} \, (\textit{id}, \textit{id})) \, \textit{fst} \cdot \textit{first}$
$\quad = \quad [\![ \; \text{products} \; ]\!]$
$\qquad \textit{dimap} \, (, ()) \, \textit{fst} \cdot \textit{first}$
$\quad = \quad [\![ \; \text{coherence of } \textit{first} \text{ with unit type: } \textit{dimap} \, \textit{fst} \, (, ()) = \textit{first} \; ]\!]$
$\qquad \textit{dimap} \, (, ()) \, \textit{fst} \cdot \textit{dimap} \, \textit{fst} \, (, ())$
$\quad = \quad [\![ \; \textit{dimap} \text{ composition} \; ]\!]$
$\qquad \textit{dimap} \, (\textit{fst} \cdot (, ())) \, (\textit{fst} \cdot (, ()))$
$\quad = \quad [\![ \; \text{products} \; ]\!]$
$\qquad \textit{dimap} \, \textit{id} \, \textit{id}$
$\quad = \quad [\![ \; \textit{dimap} \text{ identity} \; ]\!]$
$\qquad \textit{id}$

as required. ♡

**Proof (of Theorem 2).** As with adapters, one direction is quite straightforward:

$\qquad \textit{lensP2C} \, (\textit{lensC2P} \, (\textit{Lens} \, v \, u))$
$\quad = \quad [\![ \; \textit{lensC2P} \; ]\!]$
$\qquad \textit{lensP2C} \, (\textit{dimap} \, (\textit{fork} \, (v, \textit{id})) \, u \cdot \textit{first})$
$\quad = \quad [\![ \; \textit{lensP2C} \; ]\!]$
$\qquad (\textit{dimap} \, (\textit{fork} \, (v, \textit{id})) \, u \cdot \textit{first}) \, (\textit{Lens} \, \textit{id} \, \textit{fst})$

    = [[ composition ]]
    *dimap (fork (v, id)) u (first (Lens id fst))*
    = [[ *first* for *Lens* ]]
    *dimap (fork (v, id)) u (Lens fst (fork (fst · cross (id, fst), snd · snd)))*
    = [[ *dimap* for *Lens* ]]
    *Lens (fst · fork (v, id)) (u · fork (fst · cross (id, fst), snd · snd) · cross d, fork (v, id)))*
    = [[ products ]]
    *Lens v u*

For the other direction, we have:

    *lensC2P (lensP2C l) k*
    = [[ *lensP2C* ]]
    *lensC2P (l (Lens id fst)) k*
    = [[ *flip* ]]
    *flip lensC2P k (l (Lens id fst))*
    = [[ free theorem of *l*, and Lemma 9 ]]
    *l (flip lensC2P k (Lens id fst))*
    = [[ *flip* ]]
    *l (lensC2P (Lens id fst) k)*
    = [[ *lensC2P* ]]
    *l (dimap (fork (id, id)) fst (first k))*
    = [[ Lemma 10 ]]
    *l k*

so *lensC2P · lensP2C = id* as required. ♡

### C.3 Prisms

**Lemma 11.** The free theorem of *right* is that

*dimap id h k = dimap g id l* ⇒ *dimap id (plus (f, h)) (right k) = dimap (plus (f, g)) id (right l)*

□

**Lemma 12.** For given $k :: P\ A\ B$ for some co-cartesian profunctor $P$ and types $A, B$, the function *flip prismC2P k* :: *Prism A B s t* → *P s t* is a profunctor morphism from *Prism A B* to *P*. □

**Proof.** We have:

    *dimap f g (flip prismC2P k (PrismC m b))*
    = [[ *flip* ]]
    *dimap f g (prismC2P (PrismC m b) k)*
    = [[ *prismC2P* ]]
    *dimap f g (dimap m (either id b) (right k))*
    = [[ *dimap* composition ]]

**Profunctor Optics**

    $dimap\ (m \cdot f)\ (g \cdot either\ id\ b)\ (right\ k)$
$=\ [\![\ \text{sums and } either\ ]\!]$
    $dimap\ (m \cdot f)\ (either\ id\ (g \cdot b) \cdot plus\ g\ id)\ (right\ k)$
$=\ [\![\ dimap\ \text{composition}\ ]\!]$
    $dimap\ (m \cdot f)\ (either\ id\ (g \cdot b))\ (dimap\ id\ (plus\ g\ id)\ (right\ k))$
$=\ [\![\ \text{free theorem of } right\ (\text{Lemma 11}),\ \text{with } g = id,\ h = id,\ \text{and } k = l\ ]\!]$
    $dimap\ (m \cdot f)\ (either\ id\ (g \cdot b))\ (dimap\ (plus\ g\ id)\ id\ (right\ k))$
$=\ [\![\ dimap\ \text{composition}\ ]\!]$
    $dimap\ (plus\ g\ id \cdot m \cdot f)\ (either\ id\ (g \cdot b))\ (right\ k)$
$=\ [\![\ prismC2P\ ]\!]$
    $prismC2P\ (PrismC\ (plus\ g\ id \cdot m \cdot f)\ (g \cdot b))\ k$
$=\ [\![\ dimap\ \text{for } PrismC\ ]\!]$
    $prismC2P\ (dimap\ f\ g\ (PrismC\ m\ b))\ k$
$=\ [\![\ flip\ ]\!]$
    $flip\ prismC2P\ k\ (dimap\ f\ g\ (PrismC\ m\ b))$

So $dimap\ f\ g \cdot flip\ prismC2P\ k = flip\ prismC2P\ k \cdot dimap\ f\ g$ as required. ♡

We also need the following result:

**Lemma 13.**

    $dimap\ Right\ (either\ id\ id) \cdot right = id$

□

**Proof.** Writing $absurd :: 0 \to a$ for the unique function from the empty type to any other, we have:

    $dimap\ Right\ (either\ id\ id) \cdot right$
$=\ [\![\ \text{sums}\ ]\!]$
    $dimap\ (plus\ absurd\ id \cdot Right)\ (either\ id\ id) \cdot right$
$=\ [\![\ dimap\ \text{composition}\ ]\!]$
    $dimap\ Right\ (either\ id\ id) \cdot dimap\ (plus\ absurd\ id)\ id \cdot right$
$=\ [\![\ \text{free theorem of } right\ ]\!]$
    $dimap\ Right\ (either\ id\ id) \cdot dimap\ id\ (plus\ absurd\ id) \cdot right$
$=\ [\![\ dimap\ \text{composition}\ ]\!]$
    $dimap\ Right\ (either\ id\ id \cdot plus\ absurd\ id) \cdot right$
$=\ [\![\ \text{coherence of } right\ \text{with empty type: } dimap\ (either\ absurd\ id)\ Right = right\ ]\!]$
    $dimap\ Right\ (either\ id\ id \cdot plus\ absurd\ id) \cdot dimap\ (either\ absurd\ id)\ Right$
$=\ [\![\ dimap\ \text{composition}\ ]\!]$
    $dimap\ (either\ absurd\ id \cdot Right)\ (either\ id\ id \cdot plus\ absurd\ id \cdot Right)$
$=\ [\![\ \text{sums}\ ]\!]$
    $dimap\ id\ id$
$=\ [\![\ dimap\ \text{identity}\ ]\!]$
    $id$

as required. ♡

**Proof (of Theorem 3).** As with adapters, one direction is quite straightforward:

$$
\begin{aligned}
&\quad prismP2C\ (prismC2P\ (PrismC\ m\ b)) \\
&= [\![\ prismC2P\ ]\!] \\
&\quad prismP2C\ (dimap\ m\ (either\ id\ b) \cdot right) \\
&= [\![\ prismP2C\ ]\!] \\
&\quad (dimap\ m\ (either\ id\ b) \cdot right)\ (PrismC\ Right\ id) \\
&= [\![\ composition\ ]\!] \\
&\quad dimap\ m\ (either\ id\ b)\ (right\ (PrismC\ Right\ id)) \\
&= [\![\ right\ \text{for}\ PrismC\ ]\!] \\
&\quad dimap\ m\ (either\ id\ b)\ (PrismC\ (either\ (Left \cdot Left)\ (plus\ Right\ id \cdot Right))\ Right) \\
&= [\![\ dimap\ \text{for}\ PrismC\ ]\!] \\
&\quad PrismC\ (plus\ (either\ id\ b)\ id \cdot either\ (Left \cdot Left)\ (plus\ Right\ id \cdot Right) \cdot m)\ (either\ id\ b \cdot Right) \\
&= [\![\ sums\ ]\!] \\
&\quad PrismC\ m\ b
\end{aligned}
$$

For the other direction, we have:

$$
\begin{aligned}
&\quad prismC2P\ (prismP2C\ l)\ k \\
&= [\![\ prismP2C\ ]\!] \\
&\quad prismC2P\ (l\ (PrismC\ Right\ id))\ k \\
&= [\![\ flip\ ]\!] \\
&\quad flip\ prismC2P\ k\ (l\ (PrismC\ Right\ id)) \\
&= [\![\ \text{free theorem of}\ l,\ \text{and Lemma 12}\ ]\!] \\
&\quad l\ (flip\ prismC2P\ k\ (PrismC\ Right\ id)) \\
&= [\![\ flip\ ]\!] \\
&\quad l\ (prismC2P\ (PrismC\ Right\ id)\ k) \\
&= [\![\ prismP2C\ ]\!] \\
&\quad l\ (dimap\ Right\ (either\ id\ id)\ (right\ k)) \\
&= [\![\ \text{Lemma 13}\ ]\!] \\
&\quad l\ k
\end{aligned}
$$

so $prismC2P \cdot prismP2C = id$ as required. ♡

### C.4 Traversals

This proof depends on the fact that *FunList*s are traversable:

$$
\begin{aligned}
&travFunList :: Applicative\ f \Rightarrow (a \to f\ b) \to FunList\ a\ c\ t \to f\ (FunList\ b\ c\ t) \\
&travFunList\ f\ (Done\ t)\quad = pure\ (Done\ t) \\
&travFunList\ f\ (More\ x\ l) = pure\ More\ \langle * \rangle\ f\ x\ \langle * \rangle\ travFunList\ f\ l
\end{aligned}
$$

Indeed, the following lemma shows that *FunList*s are in a sense the archetypical traversable containers: an application of *traverse* in the *FunList* applicative functor can be expressed in terms of *travFunList*.

**Lemma 14.** For any $h :: A \to FunList\ B\ C\ D$, we have

**Profunctor Optics**

$\quad$ *traverse* (*Traversal h*) = *Traversal* (*travFunList h*)

$\hfill\square$

**Proof.** We prove this by structural induction over *FunList*s. Both cases start in the same way, so we capture the common reasoning first. Suppose that $h :: S \to FunList\ A\ B\ T$, and let $k = extract\ (traverse\ (Traversal\ h))$ so that *traverse* (*Traversal h*) = *Traversal k*. Then

$\quad$ *traverse* (*Traversal h*)
= $[\![$ *traverse*; *k* $]\!]$
$\quad$ *dimap out inn* (*right* (*par* (*Traversal h*) (*Traversal k*)))
= $[\![$ *par* for *Traversal* $]\!]$
$\quad$ *dimap out inn* (*right* (*Traversal* (*pair h k*)))
= $[\![$ *right* for *Traversal* $]\!]$
$\quad$ *dimap out inn* (*Traversal* (*either* (*Done* · *Left*) (*fmap Right* · *pair h k*)))
= $[\![$ *dimap* for *Traversal* $]\!]$
$\quad$ *Traversal* (*fmap inn* · *either* (*Done* · *Left*) (*fmap Right* · *pair h k*) · *out*)

so

*extract* (*traverse* (*Traversal h*)) = *fmap inn* · *either* (*Done* · *Left*) (*fmap Right* · *pair h k*) · *out*

We now prove that

$\quad$ *extract* (*traverse* (*Traversal h*)) = *travFunList h*

by structural induction over the *FunList* argument. Again, let $k = extract\ (traverse\ (Traversal\ h))$. For the base case *Done t*, we have:

$\quad$ *extract* (*traverse* (*Traversal h*)) (*Done t*)
= $[\![$ first steps (above) $]\!]$
$\quad$ *fmap inn* (*either* (*Done* · *Left*) (*fmap Right* · *pair h k*) (*out* (*Done t*)))
= $[\![$ *out* $]\!]$
$\quad$ *fmap inn* (*either* (*Done* · *Left*) (*fmap Right* · *pair h k*) (*Left t*))
= $[\![$ sums $]\!]$
$\quad$ *fmap inn* (*Done* (*Left t*))
= $[\![$ *fmap* for *FunList* $]\!]$
$\quad$ *Done* (*inn* (*Left t*))
= $[\![$ *out* $]\!]$
$\quad$ *Done* (*Done t*)
= $[\![$ *pure* for *FunList* $]\!]$
$\quad$ *pure* (*Done t*)
= $[\![$ *travFunList* $]\!]$
$\quad$ *travFunList h* (*Done t*)

For the inductive step *More x l*, we assume the inductive hypothesis

$\quad$ *extract* (*traverse* (*Traversal h*)) *l* = *travFunList h l*

and then calculate:

    *extract* (*traverse* (*Traversal h*)) (*More x l*)

= 〚 first steps (above) 〛

    *fmap inn* (*either* (*Done* · *Left*) (*fmap Right* · *pair h k*) (*out* (*More x l*)))

= 〚 *out* 〛

    *fmap inn* (*either* (*Done* · *Left*) (*fmap Right* · *pair h k*) (*Right* (*x*, *l*)))

= 〚 sums 〛

    *fmap inn* (*fmap Right* (*pair h k* (*x*, *l*)))

= 〚 *pair* 〛

    *fmap inn* (*fmap Right* (*pure* (,) ⟨∗⟩ *h x* ⟨∗⟩ *k l*))

= 〚 *inn* 〛

    *pure More* ⟨∗⟩ *h x* ⟨∗⟩ *k l*

= 〚 inductive hypothesis 〛

    *pure More* ⟨∗⟩ *h x* ⟨∗⟩ *travFunList h l*

= 〚 *travFunList* 〛

    *travFunList h* (*More x l*)

which completes the proof. ♡

Next, we show that traversal of the empty *FunList* is essentially the identity transformer, constructed from *empty* using *first* and the left unit isomorphism for products:

    *identity* :: (*Cartesian p*, *Monoidal p*) ⇒ *p a a*
    *identity* = *dimap lunit'* *lunit* (*first empty*)

**Lemma 15.**

    *dimap* (*const* (*Done t*)) *id* (*traverse k*) = *dimap id* (*const* (*Done t*)) *identity*

(where *const* :: *a* → *b* → *a* yields a constant function). □

**Proof.** We have:

    *dimap* (*const* (*Done t*)) *id* (*traverse k*)

= 〚 *traverse* 〛

    *dimap* (*const* (*Done t*)) *id* (*dimap out inn* (*right* (*par k* (*traverse k*))))

= 〚 *dimap* composition 〛

    *dimap id inn* (*dimap* (*out* · *const* (*Done t*)) *id* (*right* (*par k* (*traverse k*))))

= 〚 *out* (*Done t*) = *Left t* 〛

    *dimap id inn* (*dimap* (*const* (*Left t*)) *id* (*right* (*par k* (*traverse k*))))

= 〚 property of *right*: *dimap Left id* (*right f*) = *dimap id Left identity* 〛

    *dimap id inn* (*dimap* (*const t*) *id* (*dimap id Left identity*))

= 〚 *dimap* composition 〛

    *dimap* (*const t*) (*inn* · *Left*) *identity*

= 〚 free theorem of *identity*: *dimap f id identity* = *dimap id f identity* 〛

    *dimap id* (*inn* · *Left* · *const t*) *identity*

**Profunctor Optics**

$$= [\![ \ inn \ ]\!]$$
$$dimap\ id\ (const\ (Done\ t))\ identity$$

as required. ♡

Similarly, traversal of a singleton *FunList* is essentially a single application of the traversal body.

**Lemma 16.** The free theorem of *par* is that

$$dimap\ (cross\ (f,f'))\ (cross\ (g,g'))\ (par\ h\ h') = par\ (dimap\ f\ g\ h)\ (dimap\ f'\ g'\ h')$$

□

**Lemma 17.**

$$dimap\ single\ id\ (traverse\ k) = dimap\ id\ single\ k$$

□

**Proof.** We have:

$$dimap\ single\ id\ (traverse\ k)$$
$$= [\![ \ traverse \ ]\!]$$
$$dimap\ single\ id\ (dimap\ out\ inn\ (right\ (par\ k\ (traverse\ k))))$$
$$= [\![ \ dimap\ \text{composition} \ ]\!]$$
$$dimap\ id\ inn\ (dimap\ (out \cdot single)\ id\ (right\ (par\ k\ (traverse\ k))))$$
$$= [\![ \ out\ (single\ x) = Right\ (x, Done\ id) \ ]\!]$$
$$dimap\ id\ inn\ (dimap\ (\lambda x \to Right\ (x, Done\ id))\ id\ (right\ (par\ k\ (traverse\ k))))$$
$$= [\![ \ either\ id\ id \cdot Right = id \ ]\!]$$
$$dimap\ id\ inn\ (dimap\ (\lambda x \to Right\ (x, Done\ id))\ (either\ id\ id \cdot Right)\ (right\ (par\ k\ (traverse\ k))))$$
$$= [\![ \ dimap\ \text{composition} \ ]\!]$$
$$dimap\ id\ inn\ (dimap\ (, Done\ id)\ Right\ (dimap\ Right\ (either\ id\ id)\ (right\ (par\ k\ (traverse\ k)))))$$
$$= [\![ \ \text{Lemma 13} \ ]\!]$$
$$dimap\ id\ inn\ (dimap\ (, Done\ id)\ Right\ (par\ k\ (traverse\ k)))$$
$$= [\![ \ dimap\ \text{composition} \ ]\!]$$
$$dimap\ (, Done\ id)\ (inn \cdot Right)\ (par\ k\ (traverse\ k))$$
$$= [\![ \ \text{products},\ dimap\ \text{composition} \ ]\!]$$
$$dimap\ runit'\ (inn \cdot Right)\ (dimap\ (cross\ (id, const\ (Done\ id)))\ id\ (par\ k\ (traverse\ k)))$$
$$= [\![ \ \text{free theorem of } par\ (\text{Lemma 16}) \ ]\!]$$
$$dimap\ runit'\ (inn \cdot Right)\ (par\ k\ (dimap\ (const\ (Done\ id))\ id\ (traverse\ k)))$$
$$= [\![ \ \text{Lemma 15} \ ]\!]$$
$$dimap\ runit'\ (inn \cdot Right)\ (par\ k\ (dimap\ id\ (const\ (Done\ id))\ identity))$$
$$= [\![ \ \text{free theorem of } par\ (\text{Lemma 16})\ \text{again} \ ]\!]$$
$$dimap\ runit'\ (inn \cdot Right)\ (dimap\ id\ (cross\ (id, const\ (Done\ id)))\ (par\ k\ identity))$$
$$= [\![ \ dimap\ \text{composition} \ ]\!]$$
$$dimap\ id\ (inn \cdot Right \cdot cross\ (id, const\ (Done\ id)))\ (dimap\ runit'\ id\ (par\ k\ identity))$$
$$= [\![ \ (identity :: P\ 1\ 1) = empty\ (\text{see below}) \ ]\!]$$

$\quad$ *dimap id* (*inn* · *Right* · *cross* (*id*, *const* (*Done id*))) (*dimap runit*′ *id* (*par k empty*))
$=$ ⟦ *par* and *empty* ⟧
$\quad$ *dimap id* (*inn* · *Right* · *cross* (*id*, *const* (*Done id*))) (*dimap id runit*′ *k*)
$=$ ⟦ *dimap* composition ⟧
$\quad$ *dimap id* (*inn* · *Right* · *cross* (*id*, *const* (*Done id*)) · *runit*′) *k*
$=$ ⟦ *inn* · *Right* · *cross* (*id*, *const* (*Done id*)) · *runit*′ $=$ *single* ⟧
$\quad$ *dimap id single k*

Here, we used the fact that *identity* at the unit type is simply *empty*:

$\quad$ *identity*$_1$
$=$ ⟦ *identity* ⟧
$\quad$ *dimap lunit*′ *lunit* (*first*$_1$ *empty*)
$=$ ⟦ *first* at unit type ⟧
$\quad$ *dimap lunit*′ *lunit* (*dimap runit runit*′ *empty*)
$=$ ⟦ *dimap* composition ⟧
$\quad$ *dimap* (*lunit*′ · *runit*) (*lunit* · *runit*′) *empty*
$=$ ⟦ *lunit*′ · *runit* $=$ *id* $=$ *lunit* · *runit*′ :: 1 → 1 ⟧
$\quad$ *dimap id id empty*
$=$ ⟦ *dimap* identity ⟧
$\quad$ *empty*

where we have written *identity*$_1$ for *identity* :: *P* 1 1 and *first*$_1$ for *first* :: *P A B* → *P* (*A* × 1) (*B* × 1) as mnemonics for their more specialized types. ♡

The next lemma states that traversal of a *FunList* using a body that constructs a singleton makes a structure of singletons, and unpacking each of these singletons yields the original *FunList*. To prove this, we need the free theorem of ⟨∗⟩:

**Lemma 18.** The free theorem of ⟨∗⟩ is that

$\quad$ *fmap* (*h*·) *fs* $=$ *fmap* (·*k*) *gs* $\Rightarrow$ *fmap f* (*fs* ⟨∗⟩ *xs*) $=$ *gs* ⟨∗⟩ *fmap k xs*

□

**Lemma 19.**

$\quad$ *fmap fuse* · *travFunList single* $=$ *id*

□

**Proof.** We proceed by structural induction over the *FunList* argument. For the base case *Done t*, we have:

$\quad$ *fmap fuse* (*travFunList single* (*Done t*))
$=$ ⟦ *travFunList* ⟧
$\quad$ *fmap fuse* (*pure* (*Done t*))
$=$ ⟦ *pure* for *FunList* ⟧

**Profunctor Optics**

    *fmap fuse* (*Done* (*Done t*))
=   [[  *fmap* for *FunList*   ]]
    *Done* (*fuse* (*Done t*))
=   [[  *fuse*   ]]
    *Done t*

as required. For the inductive case *More x l*, we assume the inductive hypothesis that

    *fmap fuse* (*travFunList single l*) = *l*

and then calculate:

    *fmap fuse* (*travFunList single* (*More x l*))
=   [[  *travFunList*   ]]
    *fmap fuse* (*pure More* ⟨∗⟩ *single x* ⟨∗⟩ *travFunList single l*)
=   [[  *single*   ]]
    *fmap fuse* (*pure More* ⟨∗⟩ *More x* (*Done id*) ⟨∗⟩ *travFunList single l*)
=   [[  *pure* and ⟨∗⟩ for *FunList*   ]]
    *fmap fuse* (*More x* (*Done More*) ⟨∗⟩ *travFunList single l*)
=   [[  free theorem of ⟨∗⟩ (Lemma 18), with $k = id$   ]]
    *fmap* (*fuse*·) (*More x* (*Done More*)) ⟨∗⟩ *travFunList single l*
=   [[  *fmap* for *FunList*   ]]
    *More x* (*Done* ((*fuse*·) · *More*)) ⟨∗⟩ *travFunList single l*
=   [[  little sublemma (see below)   ]]
    *More x* (*Done* (*flip fuse*)) ⟨∗⟩ *travFunList single l*
=   [[  *More x* (*Done* (*flip f*)) ⟨∗⟩ *l* = *More x* (*fmap f l*)   ]]
    *More x* (*fmap fuse* (*travFunList single l*))
=   [[  induction   ]]
    *More x l*

which completes the proof. We used a little sublemma that

    (*fuse*·) · *More* = *flip fuse*

whose justification is simply a matter of expanding definitions.   ♡

Then we need to show that the translation from concrete to profunctor traversals is a profunctor morphism, as with the other three classes of optic.

**Lemma 20.** For given $k :: P\ A\ B$ for some cartesian, co-cartesian, monoidal profunctor $P$ and types $A, B$, the function *flip traversalC2P k* :: *Traversal A B s t* → *P s t* is a profunctor morphism from *Traversal A B* to *P*.   □

**Proof.** We have:

    *dimap f g* (*flip traversalC2P k* (*Traversal h*))
=   [[  *flip*   ]]
    *dimap f g* (*traversalC2P* (*Traversal h*) *k*)

$$\begin{aligned}
&= \quad [\![ \ \textit{traversalC2P} \ ]\!] \\
&\quad \textit{dimap f g } (\textit{dimap h fuse } (\textit{traverse k})) \\
&= \quad [\![ \ \textit{dimap composition} \ ]\!] \\
&\quad \textit{dimap } (h \cdot f) \, (g \cdot \textit{fuse}) \, (\textit{traverse k}) \\
&= \quad [\![ \ \text{naturality of } \textit{fuse} \ ]\!] \\
&\quad \textit{dimap } (h \cdot f) \, (\textit{fuse} \cdot \textit{fmap g}) \, (\textit{traverse k}) \\
&= \quad [\![ \ \textit{dimap composition} \ ]\!] \\
&\quad \textit{dimap } (h \cdot f) \, \textit{fuse} \, (\textit{dimap id } (\textit{fmap g}) \, (\textit{traverse k})) \\
&= \quad [\![ \ \text{free theorem of } \textit{traverse} \ ]\!] \\
&\quad \textit{dimap } (h \cdot f) \, \textit{fuse} \, (\textit{dimap } (\textit{fmap g}) \, \textit{id} \, (\textit{traverse k})) \\
&= \quad [\![ \ \textit{dimap composition} \ ]\!] \\
&\quad \textit{dimap } (\textit{fmap g} \cdot h \cdot f) \, \textit{fuse} \, (\textit{traverse k}) \\
&= \quad [\![ \ \textit{traversalC2P} \ ]\!] \\
&\quad \textit{traversalC2P} \, (\textit{Traversal} \, (\textit{fmap g} \cdot h \cdot f)) \, k \\
&= \quad [\![ \ \textit{dimap for Traversal} \ ]\!] \\
&\quad \textit{traversalC2P} \, (\textit{dimap f g } (\textit{Traversal h})) \, k \\
&= \quad [\![ \ \textit{flip} \ ]\!] \\
&\quad \textit{flip traversalC2P k } (\textit{dimap f g } (\textit{Traversal h}))
\end{aligned}$$

In the middle, we used a specialization of the free theorem of *traverse* to

$$\textit{dimap id } (\textit{fmap g}) \, (\textit{traverse k}) = \textit{dimap } (\textit{fmap g}) \, \textit{id} \, (\textit{traverse k})$$

Thus

$$\textit{dimap f g} \cdot \textit{flip traversalC2P k} = \textit{flip traversalC2P k} \cdot \textit{dimap f g}$$

as required. ♡

Finally, we can proceed with the proof that the concrete and profunctor representations of traversals are equivalent.

**Proof (of Theorem 4).** As with the earlier proofs, one direction is fairly straightforward:

$$\begin{aligned}
&\quad \textit{traversalP2C} \, (\textit{traversalC2P} \, (\textit{Traversal h})) \\
&= \quad [\![ \ \textit{traversalC2P} \ ]\!] \\
&\quad \textit{traversalP2C} \, (\textit{dimap h fuse} \cdot \textit{traverse}) \\
&= \quad [\![ \ \textit{traversalP2C} \ ]\!] \\
&\quad \textit{dimap h fuse } (\textit{traverse } (\textit{Traversal single})) \\
&= \quad [\![ \ \text{Lemma 14: traversal on } \textit{FunLists} \ ]\!] \\
&\quad \textit{dimap h fuse } (\textit{Traversal } (\textit{travFunList single})) \\
&= \quad [\![ \ \textit{dimap for Traversal} \ ]\!] \\
&\quad \textit{Traversal } (\textit{fmap fuse} \cdot \textit{travFunList single} \cdot h) \\
&= \quad [\![ \ \text{Lemma 19: traversal with singletons} \ ]\!] \\
&\quad \textit{Traversal h}
\end{aligned}$$

For the other direction, we have:

**Profunctor Optics**

    *traversalC2P* (*traversalP2C l*) *k*
= ⟦ *traversalP2C* ⟧
    *traversalC2P* (*l* (*Traversal single*)) *k*
= ⟦ *flip* ⟧
    *flip traversalC2P k* (*l* (*Traversal single*))
= ⟦ Lemma 20: *flip traversalC2P k* is a profunctor morphism ⟧
    *l* (*flip traversalC2P k* (*Traversal single*))
= ⟦ *flip* ⟧
    *l* (*traversalC2P* (*Traversal single*) *k*)
= ⟦ *traversalC2P* ⟧
    *l* (*dimap single fuse* (*traverse k*))
= ⟦ Lemma 17: traversal of a singleton ⟧
    *l* (*dimap id fuse* (*dimap id single k*))
= ⟦ *dimap* composition ⟧
    *l* (*dimap id* (*fuse · single*) *k*)
= ⟦ *fuse · single = id* ⟧
    *l* (*dimap id id k*)
= ⟦ *dimap* identity ⟧
    *l k*

which completes the proof.

**About the authors**

**Matthew Pickering** is a PhD student at the University of Bristol. Contact him at matthew.pickering@bristol.ac.uk.

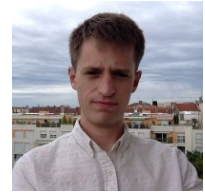

**Jeremy Gibbons** is Professor of Computing at the University of Oxford. Contact him at jeremy.gibbons@cs.ox.ac.uk.

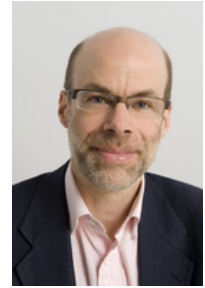

**Nicolas Wu** is a Lecturer in Computer Science at the University of Bristol. Contact him at nicolas.wu@bristol.ac.uk.

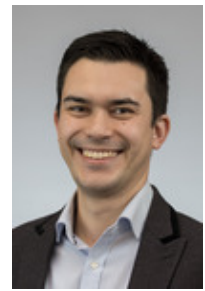